\documentclass[a4paper,11pt]{article}
\pdfoutput=1 
\usepackage{jcappub}
\usepackage{amsmath}
\usepackage{amssymb}
\usepackage{bm}
\usepackage{graphicx}
\usepackage{enumitem}
\usepackage{geometry}
\usepackage{xspace}
\usepackage{pdflscape}
\usepackage{enumerate}
\usepackage{xspace}
\usepackage{soul}
\usepackage{stmaryrd}

\makeatletter
\gdef\@fpheader{}
\g@addto@macro\bfseries{\boldmath}
\makeatother

\newcommand{\ie}{\textsl{i.e.~}}

\newcommand{\eg}{\textsl{e.g.~}}

\newcommand{\dd}{\mathrm{d}}
\newcommand{\ee}{e}

\newcommand{\sss}[1]{{\scriptscriptstyle{#1}}}
\newcommand{\boldmathsymbol}[1]{{\ensuremath{\boldsymbol{#1}}}}

\newcommand{\uPl}{\mathrm{Pl}}
\newcommand{\uin}{\mathrm{in}}

\newcommand{\umax}{\mathrm{max}}
\newcommand{\uend}{\mathrm{end}}

\newcommand{\ucl}{\mathrm{cl}}

\newcommand{\usssPl}{\sss{\uPl}}

\newcommand{\Rea}{\Re \mathrm{e}\,}
\newcommand{\Ima}{\Im \mathrm{m}\,}

\newcommand{\mpl}{m_\usssPl}
\newcommand{\Mp}{M_\usssPl}

\newcommand{\efolds}{$e$-folds}

\newcommand{\beq}{\begin{equation}}
\newcommand{\eeq}{\end{equation}}
\newcommand{\bea}{\begin{equation}\begin{aligned}}
\newcommand{\eea}{\end{aligned}\end{equation}}

\newlength{\wsingfig}
\newlength{\wdblefig}
\newlength{\wquadfig}
\newlength{\wtriplefig}
\newcommand{\Eq}[1]{Eq.~(\ref{#1})}
\newcommand{\Eqs}[1]{Eqs.~(\ref{#1})}
\newcommand{\Fig}[1]{Fig.~{\ref{#1}}}
\newcommand{\Figs}[1]{Figs.~{\ref{#1}}}
\newcommand{\Refa}[1]{Ref.~{\cite{#1}}}
\newcommand{\Refs}[1]{Refs.~{\cite{#1}}}
\newcommand{\Sec}[1]{Sec.~\ref{#1}}
\newcommand{\Secs}[1]{Secs.~\ref{#1}}
\newcommand{\App}[1]{Appendix~\ref{#1}}

\usepackage{xcolor}

\newcommand{\eq}[1]{Eq.~(\ref{#1})}
\newcommand{\fig}[1]{Fig.~\ref{#1}}

\newcommand{\uUV}{\mathrm{UV}}

\def\mpl{M_{\mathrm{Pl}}}
\def\N{\mathcal{N}}
\def\A{\mathcal{A}}
\def\B{\mathcal{B}}

\def\Nmean{\langle \mathcal{N} \rangle}

\def\d{{\mathrm d}}
\def\diffusion{S}

\def\figurewidth{0.8}
\def\halffigurewidth{0.48}
\def\gaussiancurve{solid black}

\def\kazuyacurve{solid black curve}
\def\polescurve{solid black curves}
\def\highUVcurve{dashed black curve}

\def\code{\textsc{PyFPT}\xspace}
\def\efolds{e-folds\xspace}

\def\fpt{first-passage times\xspace}

\def\pdfopening{First-passage time PDF, $P(\N)$,\xspace}

\subheader{}

\title{Numerical simulations of stochastic inflation using importance sampling}

\author[a]{Joseph H.~P. Jackson,}
\author[a,b]{Hooshyar Assadullahi,}
\author[a]{Kazuya Koyama,}
\author[c,d,a]{Vincent Vennin,}
\author[a]{David Wands}

\affiliation[a]{Institute of Cosmology \& Gravitation, University of Portsmouth, Dennis Sciama Building, Burnaby Road, Portsmouth, PO1 3FX, United Kingdom}
\affiliation[b]{School of Mathematics and Physics, University of Portsmouth, Lion Gate Building, Lion Terrace, Portsmouth, PO1 3HF, United Kingdom}
\affiliation[c]{Laboratoire de Physique de l'\'Ecole Normale Supérieure, ENS, Universit\'e PSL, CNRS, Sorbonne Universit\'e, Universit\'e Paris Cit\'e, F-75005 Paris, France}
\affiliation[d]{Laboratoire Astroparticule et Cosmologie, CNRS, Universit\'e Paris Cit\'e, 75013 Paris, France}

\emailAdd{joseph.jackson@port.ac.uk}
\emailAdd{hooshyar.assadullahi@port.ac.uk}
\emailAdd{kazuya.koyama@port.ac.uk}
\emailAdd{vincent.vennin@ens.fr}
\emailAdd{david.wands@port.ac.uk}

\date{today}

\begin{document}

\sloppy

\abstract{
We show how importance sampling can be used to reconstruct the statistics of rare cosmological fluctuations in stochastic inflation. We have developed a publicly available package, \code,\footnote{\url{https://github.com/Jacks0nJ/PyFPT}} that solves the first-passage time problem of generic one-dimensional Langevin processes. In the stochastic-$\delta N$ formalism, these are related to the curvature perturbation at the end of inflation. We apply this method to quadratic inflation, where the existence of semi-analytical results allows us to benchmark our approach. We find excellent agreement within the estimated statistical error, both in the drift- and diffusion-dominated regimes. The computation takes at most a few hours on a single CPU, and can reach probability values corresponding to less than one Hubble patch per observable universe at the end of inflation. With direct sampling, this would take more than the age of the universe to simulate even with the best current supercomputers. As an application, we study how the presence of large-field boundaries might affect the tail of the probability distribution. We also find that non-perturbative deviations from Gaussianity are not always of the simple exponential type.
}

\keywords{Statistical sampling techniques, inflation, physics of the early universe, primordial black holes 
}


\maketitle

\section{Introduction}
\label{sec_intro}

Inflation is a period of accelerated cosmic expansion in the very early universe~\cite{Starobinsky:1980te, Sato:1980yn, Guth:1980zm, Linde:1981mu, Albrecht:1982wi, Linde:1983gd}, proposed to explain its observed homogeneity, isotropy and flatness~\cite{Eisenstein_2011,Blake2011,Dawson2012,BOSS2017, Ade:2015xua, Saadeh:2016sak, Planck2018}. The process of accelerated expansion leads to microscopic quantum fluctuations in light scalar fields growing to macroscopic scales~\cite{Mukhanov:1981xt, Mukhanov:1982nu, Starobinsky:1982ee, Guth:1982ec, Hawking:1982cz, Bardeen:1983qw}. These result in primordial curvature perturbations, $\zeta$, whose imprint can be observed in the Cosmic Microwave Background (CMB)~\cite{Ade:2015xua, Ade:2015lrj, Planck2018} and which act as the seeds of cosmic structure~\cite{Lyth:2009zz}. Measurements of the CMB provide tight constraints on the dynamics of inflation on scales exiting the horizon 50-60 \efolds before the end of inflation~\cite{Liddle:2003as}. However, the properties of inflation on scales smaller than those observed in the CMB are not as strongly constrained.

While future experiments~\cite{Chluba:2015bqa} will be able to constrain inflation by observations of the stochastic gravitational wave background~\cite{Christensen:2018iqi}, a complementary probe is the possible detection of black holes formed from very large perturbations produced by inflation, known as Primordial Black Holes (PBHs)~\cite{Hawking:1971ei,Carr:1974nx,Carr:1975qj}. Not only would the detection of PBHs (or lack thereof) constrain inflation on smaller scales, but PBHs may also explain the origin of supermassive black holes~\cite{Garcia-Bellido:2017fdg}, the LIGO--Virgo--KAGRA gravitational wave detections~\cite{LIGO_VIRGO2018, LIGO_virgo2021, Abbott:2020mjq, Clesse:2020ghq, KAGRA:2022myk} and some (if not all) of the dark matter, see Refs~\cite{Carr:2016drx, Carr:2020gox, Carr:2020xqk, Green:2020jor} for recent reviews.

In the standard approach to estimate PBH abundances,  perturbation theory is used for canonical single-field inflation where the background evolution of the inflaton field is given by the Klein-Gordon equation 
\begin{equation}
    \label{eq_klein_gordon}
    \Ddot{\phi} + 3H\dot{\phi} + \frac{\d V(\phi)}{\d \phi} = 0 \, .
\end{equation}
In this expression, an over-dot denotes a time derivative, $V(\phi)$ is the field's potential and $H=\dot{a}/a$ is the Hubble rate, with $a$ the scale factor of the universe. It is related to the inflaton field and its velocity by Friedmann's equation $H^2=[V(\phi)+\dot{\phi}^2/2]/(3\Mp^2)$, with $\Mp$ being the reduced Planck mass. 

In linear perturbation theory, the probability density function (PDF) of the primordial curvature perturbation is given by a Gaussian
\begin{equation}
     \label{eq_pdf_classical}
     P(\zeta) \propto {\rm exp}\Bigg(-\frac{\zeta^2}{2 \sigma_{\zeta}^2} \Bigg) \, ,
\end{equation}
where $\sigma_{\zeta}^2$ is the variance of $\zeta$ on the scale considered. Large density perturbations, resulting from large values of $\zeta$ in the tail of the PDF, collapse to form PBHs. A criterion to form a black hole can be given in terms of the compaction function~\cite{Shibata:1999zs, Harada:2015yda} (or its smoothed version~\cite{Escriva:2019phb}), which is non-linearly related to $\zeta$~\cite{Musco:2018rwt, Young:2019yug}. A PBH forms when the compaction exceeds a threshold value, with a mass given by critical scaling~\cite{Musco:2008hv}. The production of PBHs based on the PDF given in \eq{eq_pdf_classical} is therefore expected to be suppressed by a Gaussian factor, and given that $\sigma_{\zeta}^2$ is observed to be small on CMB scales~\cite{Planck2018}, a large enhancement in $\sigma_{\zeta}^2$ on small scales is required for a significant abundance of PBHs~\cite{Carr:2020gox}. 

One mechanism for producing a peak in $\sigma_{\zeta}^2$, is a period of inflation about an inflection point in the scalar field potential driving inflation, where $\d V/ \d \phi \rightarrow 0$, resulting in the power spectrum growing rapidly~\cite{Garcia-Bellido:2017mdw, Ezquiaga:2017fvi, Germani:2017bcs, Motohashi:2017kbs, Ballesteros:2017fsr, Rasanen:2018fom, Geller:2022nkr}. However, this can also lead to quantum diffusion effects becoming non-negligible~\cite{Pattison:2017mbe, Biagetti:2018pjj, Ezquiaga:2018gbw, Firouzjahi:2018vet, Ballesteros:2020sre, Pattison:2021oen} (see \Refs{Prokopec:2019srf, Rigopoulos:2021nhv} for a different viewpoint).

The stochastic approach to inflation enables us to study non-perturbative effects where the quantum diffusion can be large. This formalism introduces a coarse-graining scale which separates short and long wavelength modes of the scalar field. Quantum field fluctuations on short wavelengths are swept up into the long wavelength regime, where they are incorporated in the coarse-grained field, $\bar{\phi}$, resulting in a stochastic noise term, $\xi$, in the dynamical equations~\cite{Starobinsky:1982ee, Starobinsky:1986fx, Nambu:1987ef, Nambu:1988je, Kandrup:1988sc, Nakao:1988yi, Nambu1989b, Mollerach:1990zf, Linde:1993xx, Starobinsky1994}. We thus model inflation as a non-perturbative stochastic process, with the results of linear perturbation theory recovered in the low-diffusion limit~\cite{Vennin:2015hra}. In the slow-roll approximation [neglecting the field acceleration in \eq{eq_klein_gordon}], this process is described by a first-order Langevin equation~\cite{Starobinsky:1986fx}
\begin{equation}
    \label{eq_langevin_sr}
    \frac{\partial \bar{\phi}}{\partial N} = -\frac{1}{3H^2(\bar{\phi})} \frac{\d V(\bar{\phi})}{\d \bar{\phi}}+\frac{H(\bar{\phi})}{2\pi}\xi \,,
\end{equation}
where the local Hubble rate $H$ is now a function of the local coarse-grained field,
\begin{equation}
    \label{eq_hubble_sr}
    H^2(\bar{\phi}) = \frac{V(\bar{\phi})}{3 \Mp^2} \,,
\end{equation}
with $\xi$ being a white Gaussian noise. As the number of \efolds elapsed during inflation, $N=\int H\d t$, has been used as the time variable, we are implicitly working in the uniform-$N$ gauge~\cite{Pattison:2019hef} and this choice allows $\zeta$ to be found by use of the $\delta N$ formalism~\cite{Starobinsky:1982ee, Starobinsky:1986fxa, Sasaki:1995aw, Sasaki:1998ug, Lyth:2004gb}. In this approach, the integrated local expansion $\N$ of a homogeneous patch, treated as a separate universe~\cite{Salopek:1990jq, Sasaki:1995aw, Wands:2000dp, Lyth:2003im, Rigopoulos:2003ak, Lyth:2005fi}, is measured from an initially flat hypersurface to a hypersurface of uniform energy density. The curvature perturbation
$\zeta$ is then given by the difference between this local expansion and its mean value,
\begin{equation}
    \label{eq_delta_N}
    \zeta =  \N - \Nmean \equiv \delta \N \,,
\end{equation}
where angle brackets $\langle \cdot \rangle$ denote the ensemble average. Finding the PDF of $\N$ thus corresponds to solving a first-passage time (FPT) problem~\cite{Vennin:2015hra}. This is the stochastic-$\delta N$ formalism~\cite{Enqvist:2008kt, Fujita:2013cna, Vennin:2015hra}, which allows $\zeta$ to be calculated beyond perturbation theory. The statistics of the curvature perturbation (and other quantities of cosmological interest such as the density contrast or the compaction function) when coarse-grained at a fixed physical scale can then be reconstructed using backward probabilities~\cite{Ando:2020fjm, Tada:2021zzj}.

If the inflating domain is bounded in field space, it has been shown that the PDF of $\N$ (and hence of $\zeta$) is given by a sum of decaying exponentials~\cite{Pattison:2017mbe}
\begin{equation}
 \label{eq_pdf_expansion_in_exponentials}
 P_{\phi}(\N) = \sum_n a_n(\phi)e^{-\Lambda_n \N} \, .
\end{equation}
Here $\phi$ denotes the initial field configuration (from now on we drop the explicit over-bar notation denoting the coarse-grained field for convenience), on which the decay rates $\Lambda_n$ do not depend. Technically, these decays rates appear as poles of the characteristic function of the PDF~\cite{Ezquiaga:2019ftu}, hence they will be refereed to as ``poles'' in what follows. Therefore, while the peak of the PDF may be well approximated by a Gaussian, the far tail is rather exponential. Beyond a very few test cases~\cite{Pattison:2017mbe, Ezquiaga:2019ftu} for which this exponential tail can be calculated analytically, numerical simulations are in general required to reconstruct the PDF of $\N$~\cite{Martin:2011ib, Kawasaki:2015ppx, De:2020hdo, Figueroa:2020jkf, Figueroa:2021zah, Mahbub:2022osb}. However, direct simulations primarily sample the peak of the distribution, corresponding to the most likely realisations, while PBHs form from those rare fluctuations living in the far tail. This implies that billions of simulations need to be run on supercomputers, from which only a tiny fraction is kept to reconstruct the tail. The poor efficiency of direct sampling therefore calls for new approaches.

In this work, we explain how the method of importance sampling~\cite{Kloek1978, Mazonka1998} can address this issue, and we apply it to stochastic inflation for the first time. Importance sampling deliberately over-samples the rare, large $\zeta$ events, which are then re-weighted to recover the true probability distribution far into the tail. We have developed the publicly available \code package,\footnote{\url{https://github.com/Jacks0nJ/PyFPT}}\ which is general and applicable to any one-dimensional Langevin equation. Here we apply it to stochastic inflation.

This paper is organised as follows. In \Sec{sec_importance_sampling} we introduce the importance sampling method and the associated data-analysis techniques. In \Sec{sec_demonstration} we use both analytical and semi-analytical test cases to illustrate the accuracy of the importance sampling method and present results for the case of slow-roll inflation driven by a scalar field with a quadratic potential. Supporting calculations for this test case are given in Appendices~\ref{sec_app_analysis} and~\ref{sec_app_edgeworth}. We draw our conclusions in \Sec{sec_conclusions}.

\section{Importance sampling}
\label{sec_importance_sampling}

\subsection{Direct sampling of a Langevin equation}
\label{sub_direct_sampling}

Consider a general one-dimensional Langevin equation
\begin{equation}
    \label{eq_langevin_basic}
    \frac{\d x}{\d t} = D(t,x)+\diffusion(t,x) \xi \,,
\end{equation}
where $x$ is the stochastic variable to be propagated, $t$ is the time variable, $D(t, x)$ is the deterministic drift and $\diffusion(t, x)$ is the amplitude of the stochastic diffusion. $\xi$ is a random white Gaussian noise, normalised such that $\langle \xi(t) \xi(t') \rangle = \delta(t-t')$. Starting from an initial state $x=x_m$ at $t=t_m$, an estimate for the stochastic variable $x=x_{m+1}$ at time $t_{m+1}=t_m+\Delta t_m$ is given, for a sufficiently small interval $\Delta t_m$, by~\cite{Kloeden1992}
\begin{equation}
    \label{eq_euler_maruyama}
    x_{m+1} = x_m + D(t_m,x_m)\Delta t_m + \diffusion(t_m,x_m)\xi_m \sqrt{\Delta t_m} \, .
\end{equation}
Here, $\xi_m$ is a random number drawn from a normal distribution with unit variance, meaning that each step is given by a Gaussian with a mean $D(t_m,x_m)\Delta t_m$ and a standard deviation $ \diffusion(t_m,x_m)\sqrt{\Delta t_m}$. This is the Euler--Maruyama method~\cite{Kloeden1992} of solving a stochastic differential equation in the It\^o prescription, which we adopt for numerical simplicity. In single-field slow-roll inflation it produces similar results to the Stratonovitch prescription, which
has been shown to preserve field-space covariance in the multiple field setup~\cite{Pinol:2020cdp}. In a first-passage time problem (FPT), the procedure \eqref{eq_euler_maruyama} is repeated until a given final condition $x_{\mathrm{end}}$ is reached, and the corresponding elapsed time $t_{\mathrm{FPT}}$ is recorded. 

To estimate the PDF for the \fpt, $P(t_{\mathrm{FPT}})$, multiple completed simulations, known as runs, are required. These runs can then be binned, with our estimate of the PDF of $j$th bin with $t_{\mathrm{FPT}}^{(j)}<t_{\mathrm{FPT}}\leq t_{\mathrm{FPT}}^{(j+1)}$ given by
\begin{equation}
    \label{eq_bin_pdf_reconstrcuted}
    \hat{P}\left(t_{\mathrm{FPT}}^{(j)}\right) = \frac{n_j}{\left(t_{\mathrm{FPT}}^{(j+1)}-t_{\mathrm{FPT}}^{(j)}\right) n_{\mathrm{total}}} \, ,
\end{equation}
where $n_{j}$ is the number of runs in the $j$th bin and $n_{\mathrm{total}} = \sum_{j}n_j$ is the total number of simulations. 

A `hat' is used to indicate that \Eq{eq_bin_pdf_reconstrcuted} is only a numerical estimate of the true distribution $P(t_{\mathrm{FPT}})$. There are indeed two sources of errors. First, since finite time steps $\Delta t_m$ are used in \Eq{eq_euler_maruyama}, each run comes with numerical error. Second, since only a finite number of runs is available, \Eq{eq_bin_pdf_reconstrcuted} is subject to statistical error, which scales as $1/\sqrt{n_{\mathrm{total}}}$ according to the central limit theorem~\cite{Araujo1980}, given that the runs are independent. By dividing the full set of runs into $n_{\mathrm{sub}}$ subsets, the estimate~\eqref{eq_bin_pdf_reconstrcuted} can be computed in each subset, and the variance of the results provides an estimate of the error when the sample size is $n_{\mathrm{total}}/n_{\mathrm{sub}}$. This can then be extrapolated to the full sample size, $n_{\mathrm{total}}$, using the scaling mentioned above. This is called jackknife resampling, which in this work we find to be reliable for $n_{\mathrm{sub}}>10$ and in bins where $n_j>400$.

As mentioned above, this direct-sampling method is efficient at reconstructing the peak of the PDF, and thus provides reliable estimates of \eg its lowest moments. However, rare events lying in the tail suffer from large statistical error.

\subsection{Importance sampling of a Langevin equation}
\label{sub_importance_sampling}

To efficiently investigate the tail of the probability distribution, a bias can be introduced into the Langevin equation (\ref{eq_langevin_basic}), to increase the occurrence of rare realisations of the stochastic process. This is done by modifying the drift term~\cite{Mazonka1998}
\begin{equation}
    \label{eq_importance_sampling_langevin_modified}
    \frac{\d x}{\d t} =[ D(t,x)+\mathcal{B}(t,x)]+\diffusion(t,x) \xi \, ,
\end{equation}
with the associated Euler--Maruyama step (\ref{eq_euler_maruyama}) becoming
\begin{equation}
    \label{eq_euler_maruyam_modified}
     x_{m+1}-x_m = [D(t_m,x_m)+\mathcal{B}(t_m,x_m)]\Delta t_m + \diffusion(t_m,x_m)\xi_m \sqrt{\Delta t_m} \, .
\end{equation}
$\mathcal{B}$ is known as the bias, as it modifies the mean of the Gaussian distribution for each step. By choosing an appropriate bias, the resulting stochastic process can primarily sample the area of interest of the probability distribution function. Importance sampling is then achieved by calculating the probability associated with the unmodified Langevin equation \eqref{eq_langevin_basic}, known as the target distribution (T), relative to the distribution given by the modified equation \eqref{eq_importance_sampling_langevin_modified}, known as the sample distribution (S). 

To calculate this relative probability, known as the weight $w$, consider a run made of $M$ numerical steps $\boldmathsymbol{X} = (x_1, x_2,...x_{M})$ starting from some initial value $x_0$ and using the modified Langevin equation (\ref{eq_importance_sampling_langevin_modified}). The statistical weight of this run is defined as
\begin{equation}
\label{eq_importance_sampling_bias}
    w(\boldmathsymbol{X}) = \frac{p_{\mathrm{T}}(\boldmathsymbol{X}|x_0)}{p_{\mathrm{S}}(\boldmathsymbol{X}|x_0)} \, ,
\end{equation}
where $p_{\mathrm{T}}$ denotes the probability that the run $\boldmathsymbol{X}$ is generated by the target stochastic process (\ref{eq_langevin_basic}), and $p_{\mathrm{S}}$ is the same probability in the sample process~\eqref{eq_importance_sampling_langevin_modified}. Since these processes are Markovian, the steps $ x_{m+1}-x_m$ are independent random variables, with Gaussian distributions since they are linearly related to $\xi_m$. The probability associated to the run $\boldmathsymbol{X}$ can thus be written as the product of the probabilities for each step,
\begin{flalign}
\label{eq_importance_sampling_probability_of_target_path} 
     p_{\mathrm{T}}(\boldmathsymbol{X}|x_0) &=  \displaystyle \prod_{m=0}^{M-1} \dfrac{1}{\diffusion(t_m, x_m)\sqrt{2\pi \Delta t_m}} \ee^ {- \dfrac{\left[x_{m+1}-x_{m}-D(t_m,x_m) \Delta t_m\right]^2}{2 \diffusion^2(t_m, x_m) \Delta t_m} } \, , &&
\end{flalign}
\begin{flalign}
\label{eq_importance_sampling_probability_of_sample_path}
    p_{\mathrm{S}}(\boldmathsymbol{X}|x_0) = \displaystyle \prod_{m=0}^{M-1} \dfrac{1}{\diffusion(t_m, x_m)\sqrt{2\pi \Delta t_m}}\ee^{- \dfrac{\left\lbrace x_{m+1}-x_{m}-[D(t_m,x_m)+\mathcal{B}(t_m,x_m)] \Delta t_m\right\rbrace ^2}{2 \diffusion^2(t_m, x_m) \Delta t_m} }  \, . &&
\end{flalign}
By substituting these expressions into \Eq{eq_importance_sampling_bias}, one obtains for the weight
\begin{equation}
\label{eq_importance_sampling_bias_w}
    w(\boldmathsymbol{X}) = \exp \bigg\{ \sum_{m=0}^{M-1}\frac{\mathcal{B}(t_m, x_m)}{\diffusion^2(t_m, x_m)}\bigg[\frac{\mathcal{B}(t_m, x_m)}{2}\Delta t_m + \diffusion(t_m, x_m) \xi_m \sqrt{\Delta t_m} \bigg] \bigg\} \, .
\end{equation}
Note that $\xi_m$ is the noise of the \textit{sample} process.

When solving \eq{eq_importance_sampling_langevin_modified}, one can then update the sum appearing in \Eq{eq_importance_sampling_bias_w} at each step. When a run $\boldmathsymbol{X}$ is complete, the weight is recorded alongside the first-passage time, and the target PDF can be estimated as 
\begin{equation}
    \label{eq_bin_pdf_reconstrcuted_weighted_naive}
    \hat{P}\left(t_{\mathrm{FPT}}^{(j)}\right) = \frac{\sum_{i=0}^{n_{j}-1} w_j^{(i)}}{\left(t_{\mathrm{FPT}}^{(j+1)}-t_{\mathrm{FPT}}^{(j)}\right)n_{\mathrm{total}}} \, ,
\end{equation}
where $w_j^{(i)}$ is weight of the $i$th run belonging to the $j$th bin. Hereafter this will be refereed to as the ``na\"ive'' estimate of the PDF. We recover the simple estimate~(\ref{eq_bin_pdf_reconstrcuted}) of the PDF for a direct simulation, where $\mathcal{B}=0$, since each run has $w_j^{(i)}=1$ that case. This method is  straightforward to implement and jackknife resampling can still be used to estimate the uncertainty. However we shall see in the following that if there is a large dispersion in the weights 
associated with different runs within a given bin, then this na\"ive method may underestimate the statistical error, which will lead us to consider alternative estimates in such cases.

\subsection{Stochastic inflation}
\label{sub_direct_sampling_stochastic_inflation}

Let us now apply importance sampling to stochastic inflation. Upon comparing \Eqs{eq_langevin_sr} and \eqref{eq_langevin_basic}, one can see that the drift and diffusion terms are given by
\begin{equation}
    D \equiv -\frac{1}{3H^2(\phi)} \frac{\d V(\phi)}{\d \phi} \, ,
    \quad\quad
    \label{eq_sr_diffusion}
    \diffusion \equiv \frac{H(\phi)}{2\pi} \, ,
\end{equation}
if time is labeled by the number of \efolds $N$ and $H^2=V/(3\Mp^2)$. To simulate a particular model of slow-roll inflation, the \code package requires the potential function, $V(\phi)$, its derivative, $\d V/\d \phi$, and the start and end values of the field, $\phi_{\mathrm{in}}$ and $\phi_{\mathrm{end}}$, as inputs.

In practice, we set the bias to only be a function of the field, which we parameterise as
\begin{equation}
    \label{eq_importance_sampling_amplitude_general}
    \mathcal{B}(\phi) = \A f(\phi) \, ,
\end{equation}
where $\A$ is the bias amplitude and $f(\phi)$ describes its field dependence. The function $f(\phi)$ can be set to optimise the convergence of the importance-sampling procedure, as we will further discuss in \Sec{sec:BiasOptimisation}. Once $f$ is fixed, varying $\mathcal{A}$ leads to sampling different regions of the distribution. Indeed, the mean number of \efolds in the sample distribution reads~\cite{2001gfpp.book.....R,Vennin:2015hra}
\begin{equation}
\label{eq:meanN:S:gen}
\left\langle \N\right\rangle_{\mathrm{S}}(\phi) =     2\int_{\phi_\uend}^\phi \dd\phi_1 \int_{\phi_1}^{\phi_{\mathrm{UV}}} \frac{\dd\phi_2}{S^2(\phi_2)}\exp\left[2\int_{\phi_1}^{\phi_2}\dd\phi_3 \frac{D(\phi_3)+\mathcal{B}(\phi_3)}{S^2(\phi_3)}\right] .
\end{equation}
Here, inflation starts from $\phi$ and is allowed to take place between $\phi_\uend$ (where it ends) and $\phi_{\mathrm{UV}}$ (which may be infinite and where a reflective boundary is placed~~\cite{Assadullahi:2016gkk, Vennin:2016wnk}). By varying $\mathcal{A}$, one can tune $\langle \N \rangle_{\mathrm{S}}$, hence one can choose the typical values of $\N$ that are best sampled. In practice, until we further discuss how the $f$ function can be optimised, we take it to match the noise amplitude, \ie $f(\phi)= S(\phi)$.

\subsection{Lognormal estimator}
\label{sub_lognormal_estimator}

\begin{figure}
\begin{center}
        \includegraphics[width=\halffigurewidth\textwidth]{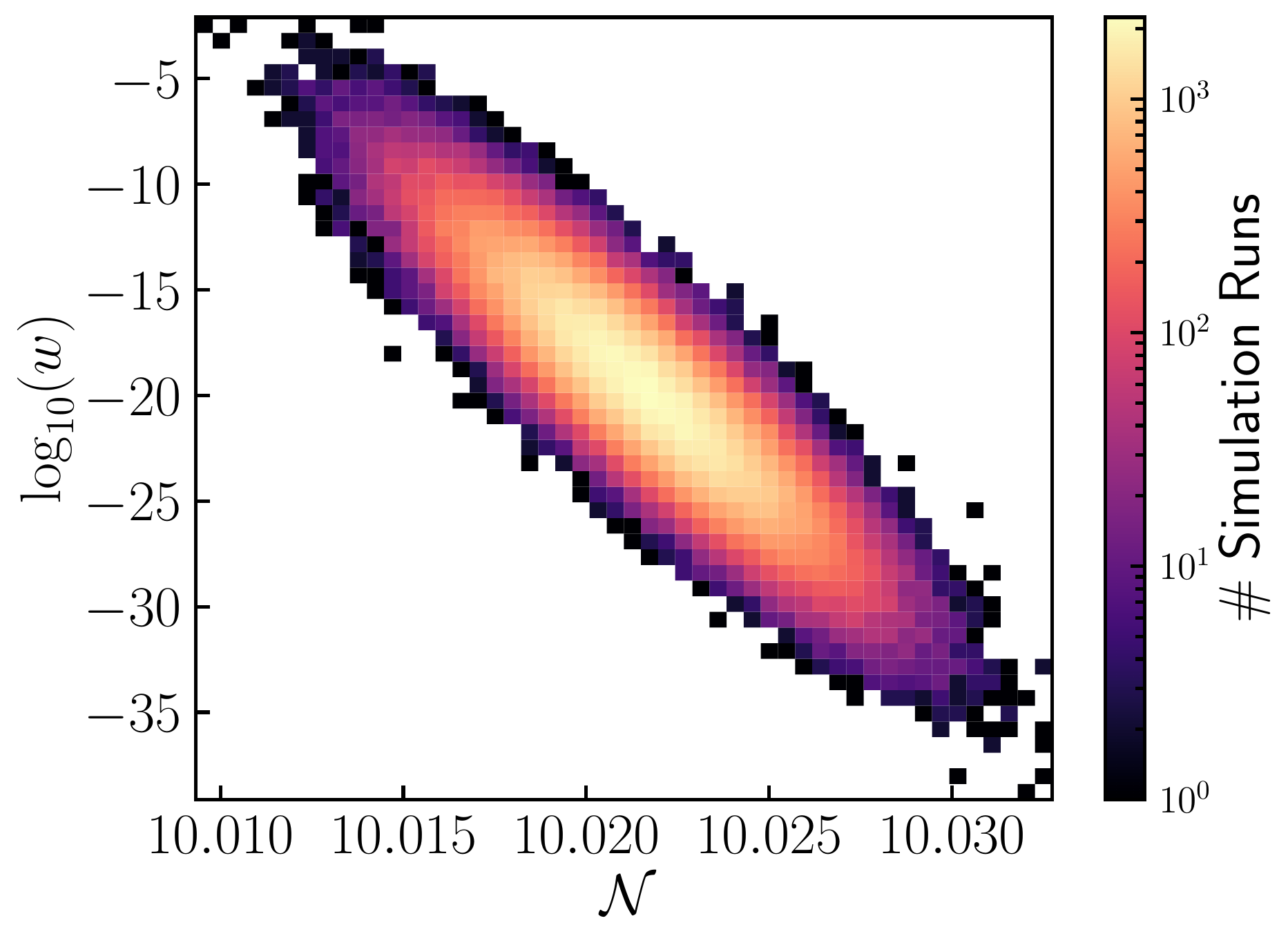}
        \includegraphics[width=\halffigurewidth\textwidth]{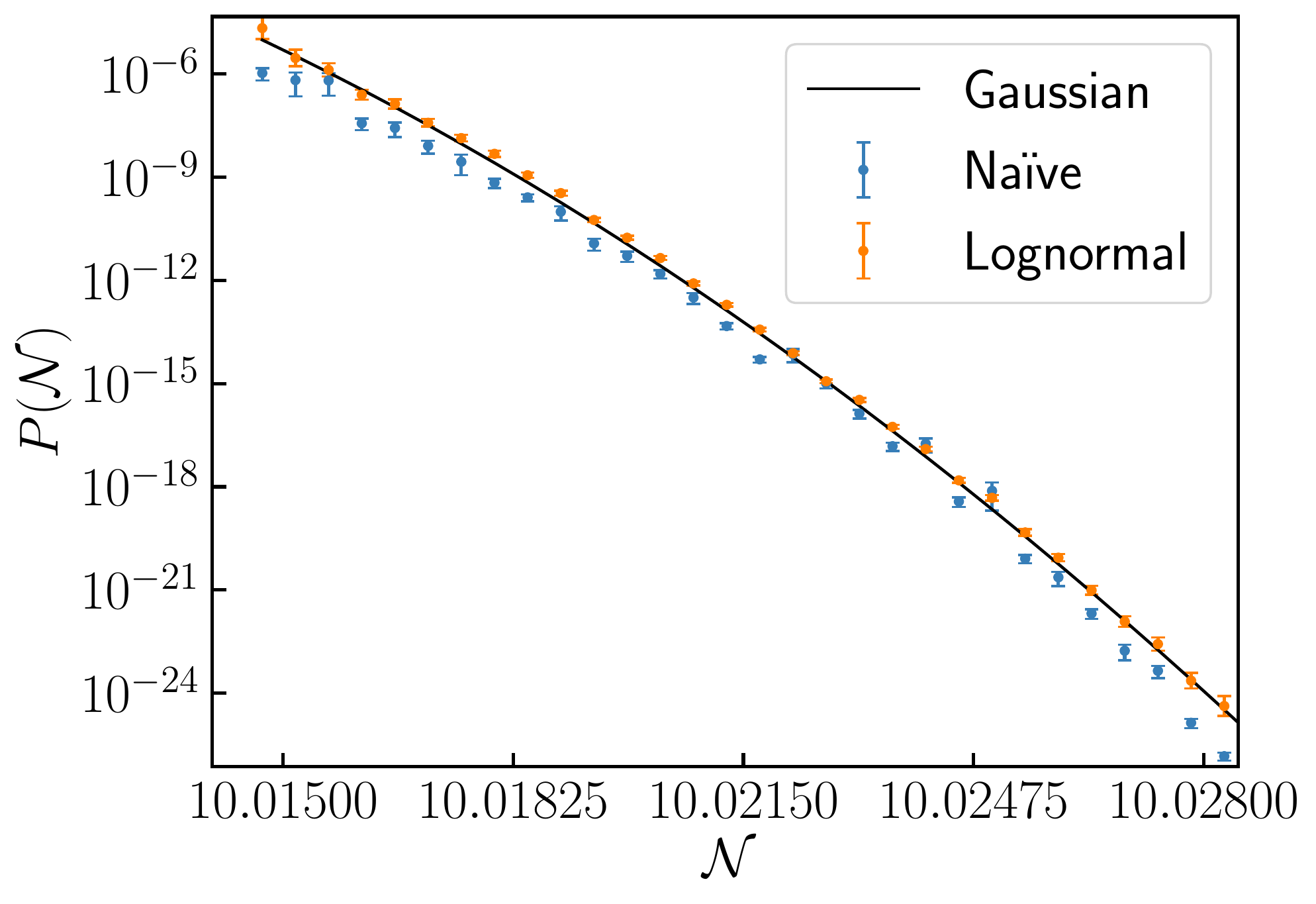}
        \caption{Left panel: a 2D histogram of the logarithm of the weights, $\log_{10}(w)$, with \fpt, $\N$, for an importance sampling simulation of $2 \times 10^5$ runs for quadratic inflation (\ref{eq_quadratic_inflation_potential}) with mass $m=0.001\Mp$, initial field value $\phi_{\uin}= \sqrt{42}\Mp$ and bias amplitude $\A=3$. 
        Right panel: estimations of the PDF, $\hat{P}(\N)$, using this data, with the na\"ive method (\ref{eq_bin_pdf_reconstrcuted_weighted_naive}), shown in blue and the lognormal method (\ref{eq_lognormal_bin_pdf_reconstrcuted_weighted_maximum_likelihood}), shown in orange. The analytical prediction for $P(\N)$ in the low-diffusion limit, corresponding to a Gaussian distribution (see \App{sec_app_edgeworth}), is shown as a \gaussiancurve~curve.}
        \label{fig_showing_scatter_in_w_and_error}
\end{center}
\end{figure}

The value of the weights within the $j$th bin, $w_j^{(i)}$, can vary by several orders of magnitude when sampling the far tail of $P(\N)$, as illustrated in the left-hand panel of \fig{fig_showing_scatter_in_w_and_error}, for the model that will be considered in \Sec{sec_demonstration}.
In this case, a very large number of simulation runs are needed to prevent the na\"ive estimator (\ref{eq_bin_pdf_reconstrcuted_weighted_naive}) from being dominated by only a small number of runs with the highest weights, leading to a systematic error in $\hat{P}(\N)$. This greatly reduces the numerical efficiency. 

To illustrate this problem, consider the distribution of weights for the bin centred at $\N=10.02$ in \fig{fig_showing_scatter_in_w_and_error} for quadratic inflation (\ref{eq_quadratic_inflation_potential}) with mass $m=0.001\Mp$, initial field value $\phi_{\uin}= \sqrt{42}\Mp$ and bias amplitude $\A=3$. As the value of $w$ varies by 8 orders of magnitude in this bin, only the runs with the largest, but rare, $w$ values dominate the na\"ive estimate of the PDF in \eq{eq_bin_pdf_reconstrcuted_weighted_naive}. The majority of the runs contribute very little as they have much smaller weights. If by chance a few more, very large $w$ values are sampled, then an over estimation occurs. Equally, if very few of the rare, but large $w$ values are randomly sampled, then an underestimation occurs. The jackknife resampling used to estimate the uncertainty in the estimate, $\hat{P}(\N)$, also suffers a systematic error for similar reasons. This effect can be seen in the right-hand panel of \fig{fig_showing_scatter_in_w_and_error}, where most points obtained using the naive estimator lie below the analytical prediction, and just two points lie above the line.

In practice, one requires at least $n> \langle w \rangle ^2 \exp ({\sigma_{\ln w}^2})$ runs to overcome this effect~\cite{Shen2006}, where $\sigma_{\ln w}$ is the standard deviation of  $\ln w$. Therefore, for the simulation shown in \fig{fig_showing_scatter_in_w_and_error}, which already contains $2\times 10^5$ runs, an order of magnitude more simulations are required for the na\"ive estimator $\hat{P}(\N)$ to converge on the true $P(\N)$.

The situation can be improved provided the distribution function of the weights is known, or at least can be approximated. In the present case, with the choice $f(\phi)=S(\phi)$, \Eq{eq_importance_sampling_bias_w} leads to 
\begin{equation}
w(\boldmathsymbol{X})=\exp\left(\frac{\mathcal{A}^2}{2}\sum_{m=0}^{M-1}\Delta t_m + \mathcal{A}\sum_{m=0}^{M-1}\sqrt{\Delta t_m}\xi_m \right)
=\exp\left(\frac{\mathcal{A}^2}{2} \mathcal{N}_{\mathrm{S}} + \mathcal{A} \sqrt{\mathcal{N}_{\mathrm{S}}} \xi \right),
\end{equation}
where $\xi$ is a random normal variable with vanishing mean and unit variance, independent of $\N_{\mathrm{S}}$, namely the duration of inflation in the sample process. This leads to the following distribution for the weight
\begin{equation}
\label{eq:P:w}
    P(w)=\frac{1}{w \sqrt{2\pi}} \int_0^\infty \dd\N_{\mathrm{S}} \frac{P(\N_{\mathrm{S}})}{\mathcal{A}\sqrt{\N_{\mathrm{S}}}}\exp\left[-\frac{\left(\ln w-\frac{\mathcal{A}^2 \N_{\mathrm{S}}}{2}\right)^2}{2 \mathcal{A}^2\N_{\mathrm{S}}}\right] .
\end{equation}
This relates the PDF of $w$ to the PDF of $\N$, which is precisely the object we are trying to reconstruct. However, it can serve as the basis of an iterative procedure. Indeed, in the limit where diffusion is sub-dominant, $P(\N_{\mathrm{S}})$ is a Dirac distribution centered on its classical value $\N_{\mathrm{S,cl}}=\int_{\phi}^{\phi_\uend}\dd \phi'/[ D(\phi')+\mathcal{B}(\phi') ]$, which can be obtained by either setting the noise to zero in \Eq{eq_importance_sampling_langevin_modified} or by performing a saddle-point expansion of \Eq{eq_importance_sampling_amplitude_general} in the same limit. In that case, $P(w)$ is nothing but a lognormal distribution, \ie $\ln w $ is normally distributed with $\langle \ln w \rangle= \mathcal{A}^2 \N_{\mathrm{S,cl}}/2$ and $\sigma_{\ln w}=\mathcal{A}^2 \N_{\mathrm{S,cl}}$. 

This result only applies to the low-diffusion limit, but it may be used as a starting point to reconstruct $P(\N_{\mathrm{S}})$, which can then be updated in \Eq{eq:P:w}, leading to a new estimate for $P(w)$, so on and so forth.

In practice, \code first establishes whether or not $w$ is lognormally distributed, by applying D'Agostino and Pearson's normality test~\cite{DAgostino1973} to $\ln (w)$. This method gives the probability, or $p$-value, that a sample is drawn from an underlying normal distribution. If any $p$-values are smaller than a specified threshold (in practice we shall use 0.5\%), the \code package identifies that $w$ does not follow a lognormal distribution and alerts the user. If, however, the $p$-values are all greater than the threshold, then $P(w)$ is assumed to be lognormal. This implies a relationship between the mean of $w$ and the mean of its logarithm, namely
\begin{equation}
\label{eq_lognormal_mean_maximum_likelihood}
\langle w\rangle = \exp\left(\langle \ln w\rangle + \frac{\sigma_{\ln w}^2}{2}\right) .   
\end{equation}
Crucially, the statistical reconstruction of $\langle \ln w\rangle$ is more robust than the one of $\langle w \rangle$. Indeed, in the later, one is dominated by those few occurrences that have a large value of $w$, while all runs with negligible values of $w$ are indistinguishable from $w=0$ when evaluating $\widehat{\langle w_j\rangle} = \sum_{i=0}^{n_j-1} w_j^{(i)}$. This is the sampling problem mentioned above, when $w_j^{(i)}$ spans several orders of magnitude. This is however not an issue for $\langle \ln w \rangle$, since $\ln w_j^{(i)}$ does not cover several orders of magnitude. The strategy is therefore the following~\cite{Shen2006}: estimate $\langle \ln w_j\rangle$ and $\sigma_{\ln w_j}$ from performing ensemble averages in the set of $\ln w_j^{(i)}$ values, and deduce $\langle w_j \rangle$ from \Eq{eq_lognormal_mean_maximum_likelihood}. One can then replace \Eq{eq_bin_pdf_reconstrcuted_weighted_naive} with
\begin{equation}
    \label{eq_lognormal_bin_pdf_reconstrcuted_weighted_maximum_likelihood}
     \hat{P}(\N_j) = \frac{n_j \widehat{\langle w_j \rangle}}{n_{\mathrm{total}} (\N_{j+1}-\N_{j})} \, .
\end{equation}
Likewise, the uncertainty in $P(\N_j)=\hat{P}(\N_j)\pm \hat{\epsilon}_j^{\pm}$ is given by the standard error in estimating $\langle \ln w_j \rangle + \sigma_{\ln w_j}^2/2$~\cite{Zhou1997}, propagated through the exponential of \eq{eq_lognormal_mean_maximum_likelihood}
\begin{equation}
    \label{eq_log_normal_bin_pdf_reconstrcuted_error_maximum_likelihood}
     \hat{\epsilon}_{j}^{\pm} = \frac{n_j \widehat{\langle w_j\rangle}}{n_{\mathrm{total}} (\N_{j+1}-\N_{j})} \Bigg|\exp \bigg[ \pm \sqrt{ \frac{(\hat{\sigma}_{\ln w_j})^2}{n_j} +  \frac{(\hat{\sigma}_{\ln w_j})^4}{2n_j-2} }  \bigg] -1\bigg| \, .
\end{equation}

Using the lognormal estimators~\eqref{eq_lognormal_bin_pdf_reconstrcuted_weighted_maximum_likelihood} and~\eqref{eq_log_normal_bin_pdf_reconstrcuted_error_maximum_likelihood} leads to more robust estimates in the low-diffusion regime. It greatly reduces the number of simulation runs required to obtain a specified accuracy, as $\hat{P}(\N_j)$ now benefits from all the runs in the $j$th bin, rather than just the few ones with the largest weights $w_j^{(i)}$. It also prevents the systematic error in the estimation of ${P}(\N)$ from using the na\"ive estimate, as well as improves the uncertainty estimation, as can be checked explicitly in the right panel of \fig{fig_showing_scatter_in_w_and_error}.

\section{Quadratic inflation}
\label{sec_demonstration}

\begin{figure}[t]
\begin{center}
        \includegraphics[width=\figurewidth\textwidth]{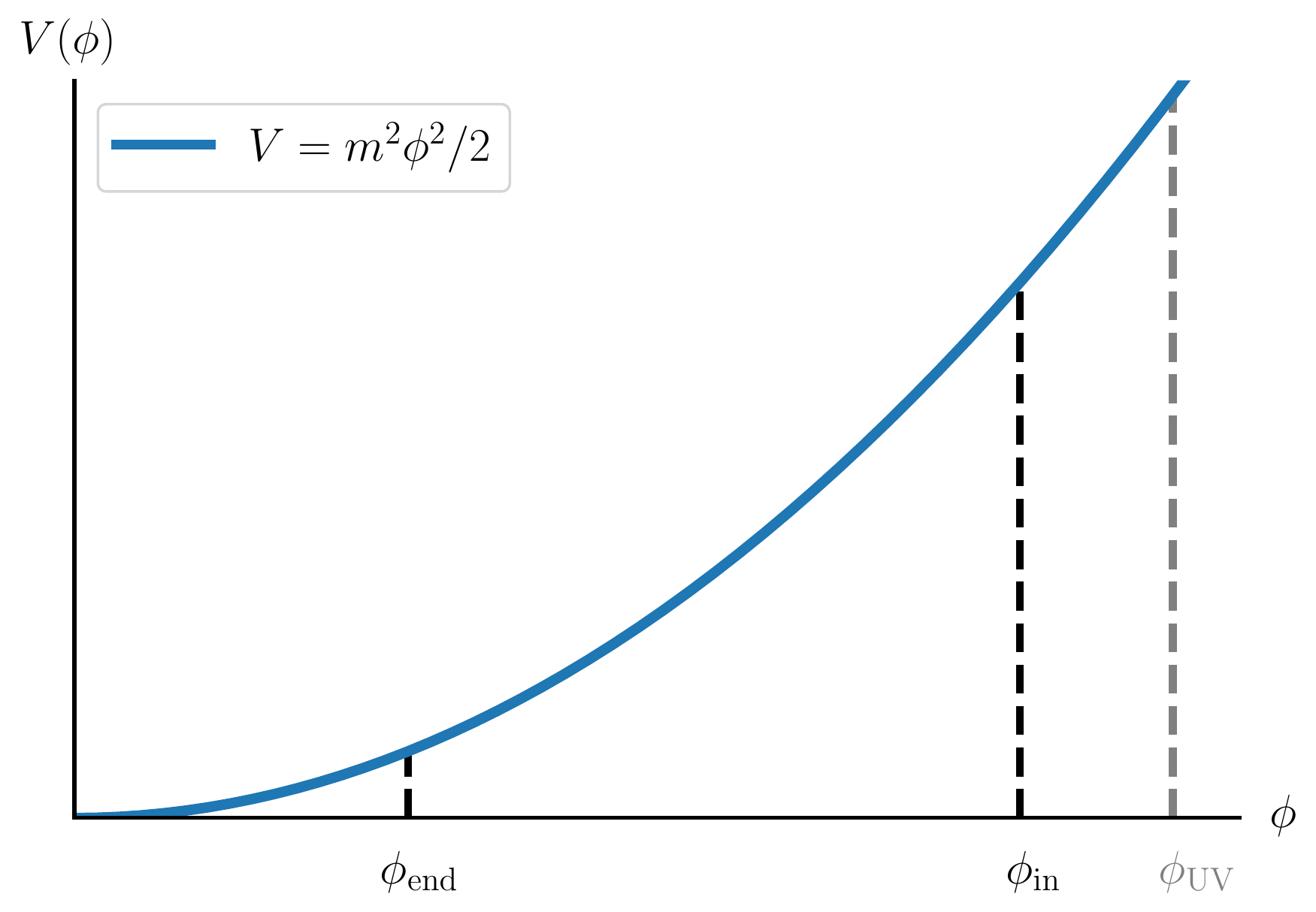}
        \caption{
        Sketch of the potential (\ref{eq_quadratic_inflation_potential}) for quadratic inflation. The initial value of the field is given by $\phi_\uin>\phi_{\uend}$ and inflation ends at $\phi_{\uend} = \sqrt{2}\Mp$, when the first slow-roll parameter becomes large, \ie when $\epsilon_1 = -\dot{H}/H^2 = 1$. We introduce a high-energy UV reflective boundary at $\phi_{\mathrm{UV}}$, shown in dashed grey.
        }
        \label{fig_quadratic_inflation}
\end{center}
\end{figure}

To test our implementation of importance sampling using the \code package, we will compare our numerical method against analytical results. For stochastic slow-roll inflation, only a few analytic solutions of the first-passage time problem are known in the literature, among which are the quantum well, with a constant potential over a finite interval, and inflation driven by a quadratic potential~\cite{Pattison:2017mbe}. The quantum well is dominated by quantum diffusion at every stage so the benefit in using importance sampling is not as striking as in quadratic inflation, on which we will thus focus.

The scalar field potential for quadratic inflation is sketched in \fig{fig_quadratic_inflation} and reads
\begin{equation}
    \label{eq_quadratic_inflation_potential}
    V(\phi) = \frac{m^2 \phi^2}{2} \, .
\end{equation}
In the last few \efolds of inflation, in which we will be mainly interested, three regimes can distinguished: diffusion dominates over the drift when $m \gtrsim\Mp$, there is an intermediate regime when $m \lesssim \Mp$, and the drift dominates when $m \ll \Mp$. These three cases will be investigated in \Secs{sub_diffusion_dom}, \ref{sub_interim_case} and  \ref{sub_drift_dom} respectively. 

Although quadratic inflation driven by the simple potential \eqref{eq_quadratic_inflation_potential} is not consistent with CMB anisotropies observed on large scales~\cite{Planck2018}, it may still describe the last \efolds of inflation, when the inflaton approaches a simple minimum of the potential with effective mass $m$ at the minimum. The runs are therefore started at an initial field value $\phi_{\uin} = \sqrt{42} \Mp$, which corresponds to $\N_\ucl=10$ \efolds before the end of inflation in the  drift-dominated limit. Obviously, values of $m$ close to or larger than the Planck mass are not allowed (since they would imply that inflation proceeds at super-Planckian energies), so here we consider the cases $m \gtrsim\Mp$ and $m \lesssim \Mp$ only to test our code in regimes where stochastic noise dominates. Indeed, there are models for which quantum diffusion dominates even at sub-Planckian energies (see \eg \Refs{Ezquiaga:2018gbw, Figueroa:2021zah}), for which this analysis is relevant. Inflation ends by slow-roll violation once the field reaches $\phi_{\uend} = \sqrt{2}\Mp$ and the \fpt are recorded. Semi-analytical results for the resulting PDF for \fpt that will be used as benchmarks for the code are presented in Appendix~\ref{sec_app_analysis}. 

Stochastic diffusion can result in the field climbing arbitrarily far back up the potential. In order to prevent the field from exploring too far up the potential, one may introduce a reflecting boundary at a finite field value, $\phi_{\mathrm{UV}}$. Formally, this boundary can be removed by sending the parameter $\phi_{\mathrm{UV}}$ to infinity, which in the case of single-field quadratic inflation does not lead to any divergence in the FPT statistics~\cite{Assadullahi:2016gkk,Vennin:2016wnk}. In the following we will numerically investigate the dependence of PDF of \fpt 
on $\phi_{\mathrm{UV}}$.

To emphasise the efficiency of \code, only $2\times 10^5$ runs will be used for each simulation set shown. The longest simulation time, required for drift-dominated cases, was $\sim8$ hours using a laptop with a quad-core processor at 1.8\,GHz base frequency. The longer simulation time for the sub-dominant diffusion case is due to the requirement that the step size $\Delta N$ must be much smaller than the standard deviation of $\N$. For diffusion-dominated and interim cases however, the simulations only took $\sim$2 minutes and $\sim$8 minutes respectively.

\subsection{Diffusion domination: $m \gtrsim \Mp$}
\label{sub_diffusion_dom}
\begin{figure}
\begin{center}
        \includegraphics[width=\figurewidth\textwidth]{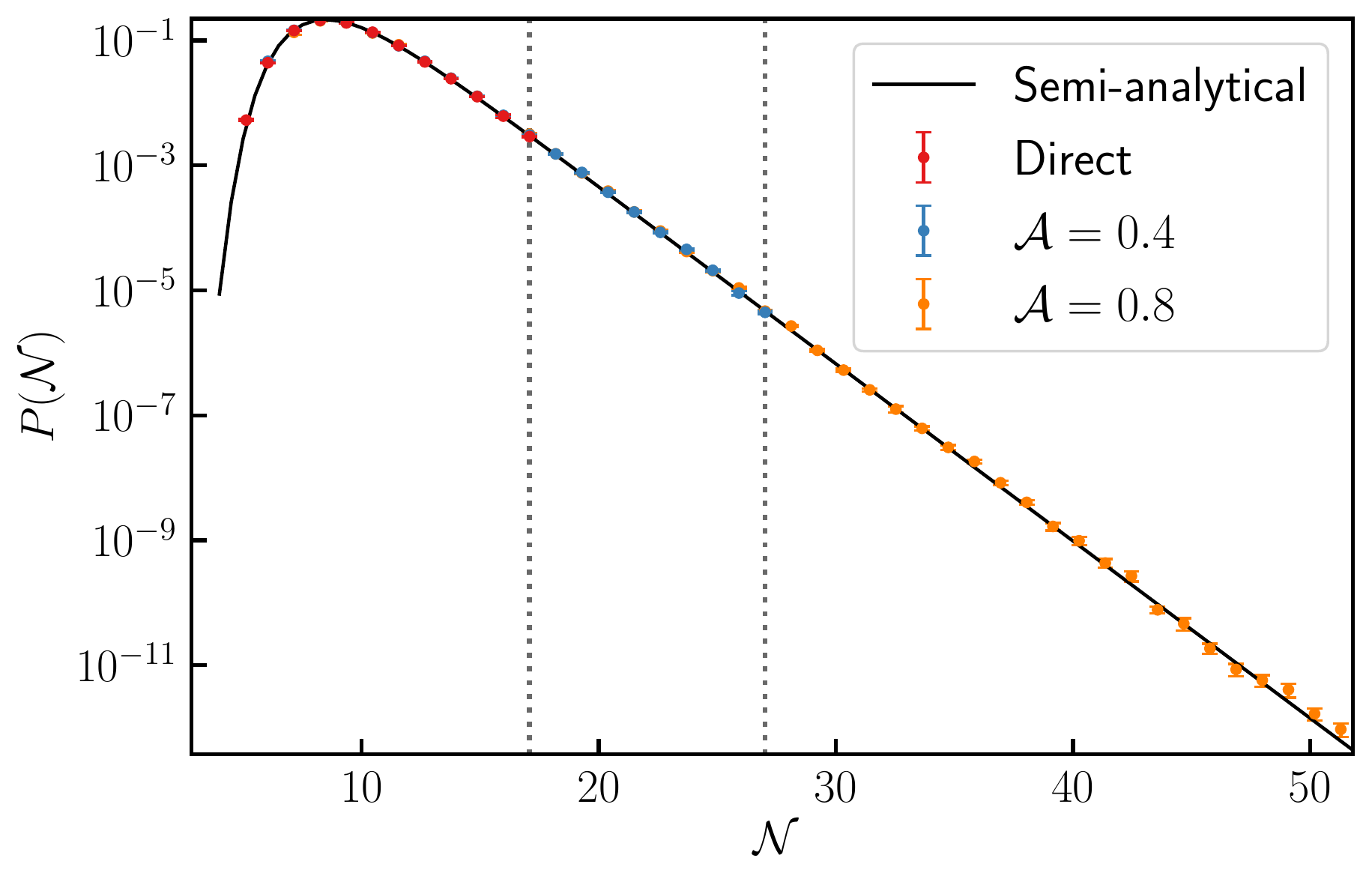}
        \caption{\pdfopening for quadratic inflation (\ref{eq_quadratic_inflation_potential}) with $m=\Mp$ and $\phi_{\mathrm{UV}} = \phi_{\uin}= \sqrt{42}\Mp$. The red, blue and orange data points correspond to bias amplitudes $\A=0$ (direct sampling), $\A=0.4$ and $\A=0.8$ respectively. The \kazuyacurve~is given by numerically calculating \eq{eq_appendix_pdf_for_residuals} for the first 50 poles. The dotted gray lines show the maximum $\N$ achieved for the direct and $\A=0.4$ data sets.}
        \label{fig_large_mass_case_finite_phi_UV}
\end{center}
\end{figure}

As a first test of the importance sampling method we investigate the diffusion-dominated regime, $m \gtrsim \Mp$. The probability of substantial deviations from the mean first-passage time is large for this case, therefore allowing us to investigate large deviations from a Gaussian distribution in the tail of the PDF, and in particular exponential tails in the presence of a finite boundary at $\phi_{\mathrm{UV}}$~\cite{Ezquiaga:2019ftu}. As will be further discussed below, the na\"ive estimator (\ref{eq_bin_pdf_reconstrcuted_weighted_naive}) is used in the figures of this subsection.

In \fig{fig_large_mass_case_finite_phi_UV} we show the results of numerical simulations of stochastic inflation for the \fpt distribution $P(\N)$, when $m=\Mp$ and $\phi_\uin = \phi_{\mathrm{UV}}$. For a direct simulation using $2\times10^5$ runs we are only able to probe the PDF down to $P(\N)\sim 10^{-3}$. Beyond this point there are too few runs in each bin ($n_j<400$) to be able to reliably estimate the statistical error using jackknife resampling. However by introducing a bias we are able to map the PDF down to $P(\N)\sim 10^{-6}$ with a bias amplitude of $\A=0.4$, while data for $\A=0.8$ reach down to $P(\N)\sim 10^{-12}$, using the same total number of runs ($2\times10^5$) in each case. This shows the numerical efficiency of the importance sampling method, and that as the bias amplitude $\A$ is increased, one probes further into the tail, up to a numerical limit discussed later.

As expected, all three simulations show large deviations from a Gaussian distribution and an exponential tail for $P(\N)$. The numerical results accurately reproduce the semi-analytical curve shown in \fig{fig_large_mass_case_finite_phi_UV}, which corresponds to keeping the first 50 poles in \Eq{eq_pdf_expansion_in_exponentials}, see \App{sub_app_numerical_reconstruction}. For $10\ll\N\leq 50$, it turns out to be dominated by the next-to-leading order pole $\Lambda_1$, and is well-described by a simple exponential tail. The leading-order pole $\Lambda_0$ is seen further into the tail at $\N>60$, beyond the regime that we can probe with  $2\times 10^5$ runs. 

\begin{figure}
\begin{center}
        \includegraphics[width=\figurewidth\textwidth]{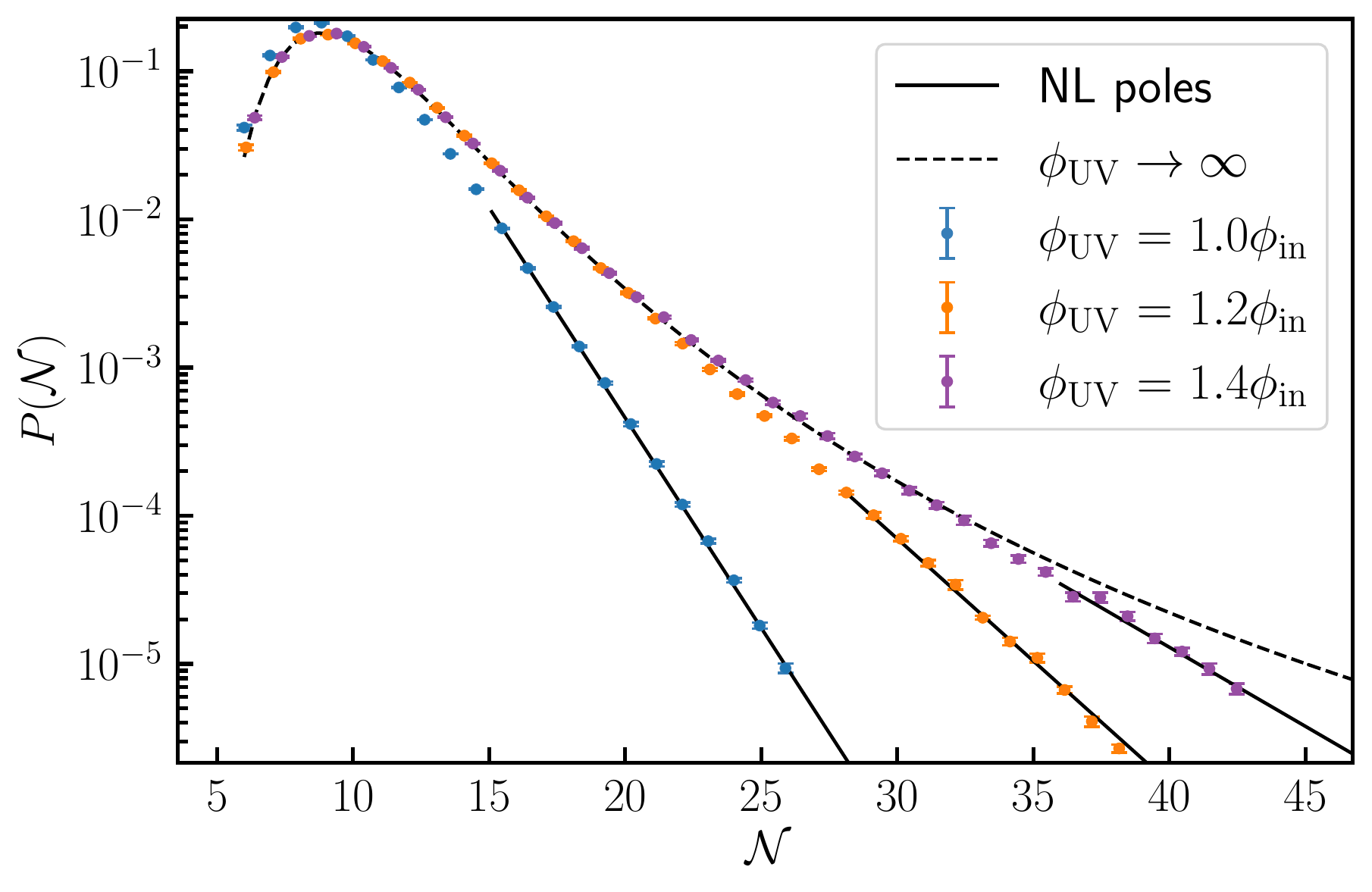}
        \caption{\pdfopening for different locations of the reflective boundary $\phi_{\mathrm{UV}}$, in quadratic inflation (\ref{eq_quadratic_inflation_potential}) with $m=\Mp$, $\phi_{\uin}= \sqrt{42}\Mp$ and bias amplitude $\A = 0.4$. The semi-analytical, \polescurve~only contains the next-to-leading pole $\Lambda_1$ in \Eq{eq_pdf_expansion_in_exponentials}, and the \highUVcurve~corresponds to the limit $\phi_{\mathrm{UV}}\rightarrow \infty$, which is obtained by numerically evaluating \Eq{eq_appendix_pdf_infinite_UV}.}
        \label{fig_large_mass_case_phi_UV_comparisons}
\end{center}
\end{figure}

In \fig{fig_large_mass_case_phi_UV_comparisons} we show the effect of varying the position of the reflecting boundary $\phi_{\mathrm{UV}}$. Analytically, we find that the distance between the poles in \Eq{eq_pdf_expansion_in_exponentials} decreases as $ \ln^{-2}(\phi_{\mathrm{UV}})$, while the leading pole $\Lambda_0$ is independent of $\phi_{\mathrm{UV}}$ (see Appendix \ref{sub_app_pole_distance}). This implies that the exponential tail, \ie the regime dominated by a single pole, occurs further into the tail with increasing $\phi_{\mathrm{UV}}$. This is exactly what is seen in \fig{fig_large_mass_case_phi_UV_comparisons}, where our numerical results agree, within the estimated error, with the expected exponential tail (where the first two poles are kept) for sufficiently large $\N$. For $\phi_{\mathrm{UV}}=1.2\,\phi_{\uin}$ and $\phi_{\mathrm{UV}}=1.4\,\phi_{\uin}$, one can also see that the neighbourhood of the peak of the distribution is well reproduced by the $\phi_{\mathrm{UV}} \rightarrow \infty$ limit, see \App{sub_app_infinite_phi_UV}, and the degeneracy between different values of $\phi_{\mathrm{UV}}$ is only lifted far in the tail. This is because the reflective boundary only affects the runs that are reflected against it, \ie those that are sufficiently long to explore the large-field regions of the potential.

\begin{figure}
\begin{center}
        \includegraphics[width=\halffigurewidth\textwidth]{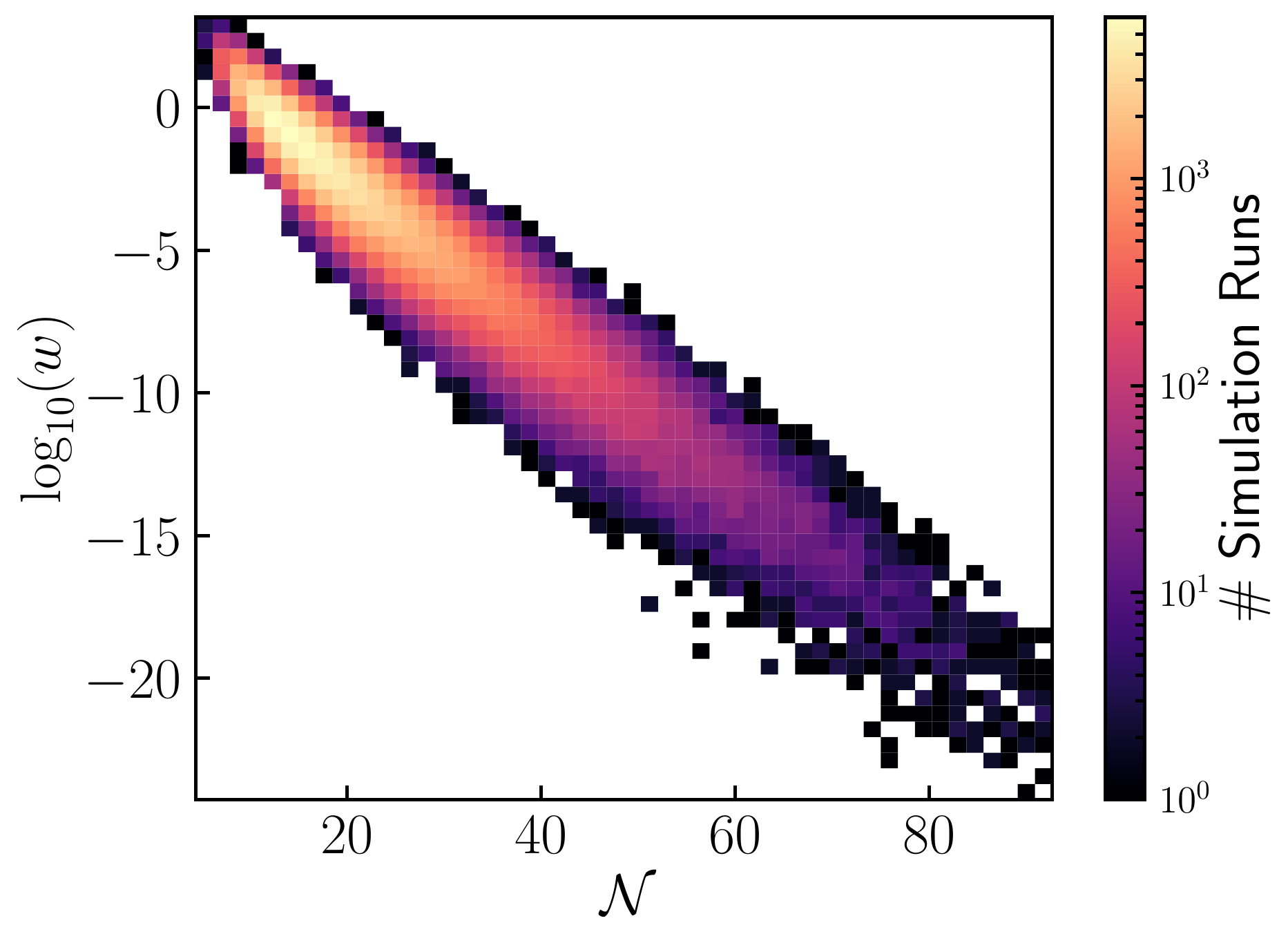}
        \includegraphics[width=\halffigurewidth\textwidth]{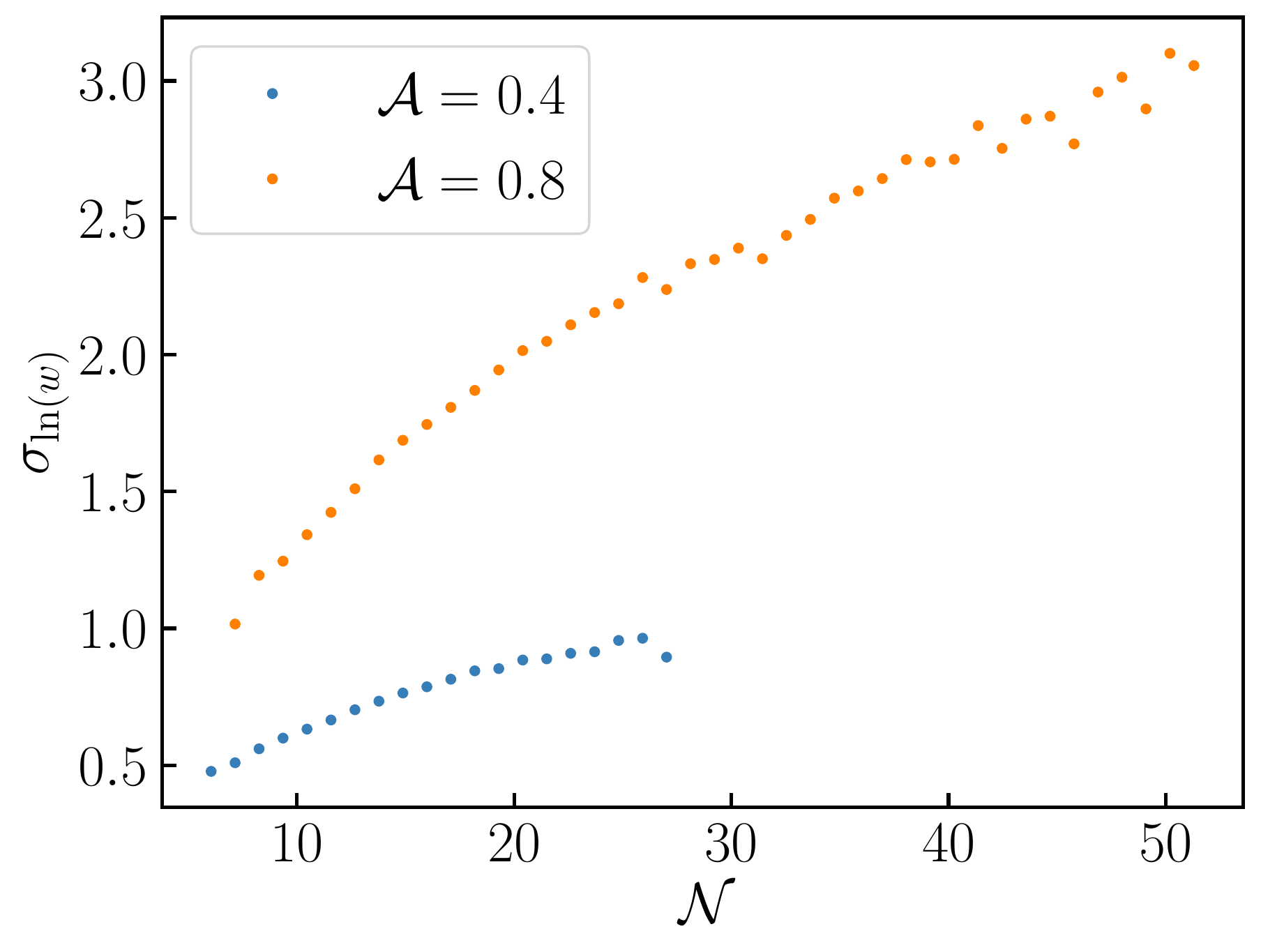}
        \caption{Left panel: a 2D histogram of the logarithm of the weights, $\log_{10}(w)$, with \fpt $\N$ for the $\A=0.8$ data shown in \fig{fig_large_mass_case_finite_phi_UV} before truncation of the bins with less than 400 runs. Right panel: the standard deviation of $\ln (w)$ in each of the $\N$ bins after truncation for the two importance samples shown in \fig{fig_large_mass_case_finite_phi_UV}.}
        \label{fig_large_mass_case_finite_phi_UV_scatter}
\end{center}
\end{figure}

Note that we find the D'Agostino and Pearson's normality test to always fail in this regime $m \gtrsim \Mp$, where almost all the $p$-values are found to be below the 0.5\% threshold. This means that the weights do not have a lognormal distribution, as illustrated in the left panel of \fig{fig_large_mass_case_finite_phi_UV_scatter}. This is because quantum diffusion is not sub-dominant in this case, and it implies that the na\"ive estimator \eqref{eq_bin_pdf_reconstrcuted_weighted_naive} for reconstructing the \fpt PDF should be used, see \Sec{sub_lognormal_estimator}.

Moreover, we find that the variance of the logarithms of the weights, $\ln (w)$, increases with $\N$ and with the bias amplitude $\A$, as shown in the right panel of \fig{fig_large_mass_case_finite_phi_UV_scatter}. Therefore, as $\A$ is increased to probe further into the tail of $P(\N)$, the accuracy of the na\"ive estimator~\eqref{eq_bin_pdf_reconstrcuted_weighted_naive} decreases, as discussed in \Sec{sub_lognormal_estimator}. This explains why this method does not allow us to reconstruct arbitrarily far regions of the tail with a fixed number of runs.

\subsection{Interim case: $m \lesssim \Mp$}
\label{sub_interim_case}
\begin{figure}
\begin{center}
        \includegraphics[width=\figurewidth\textwidth]{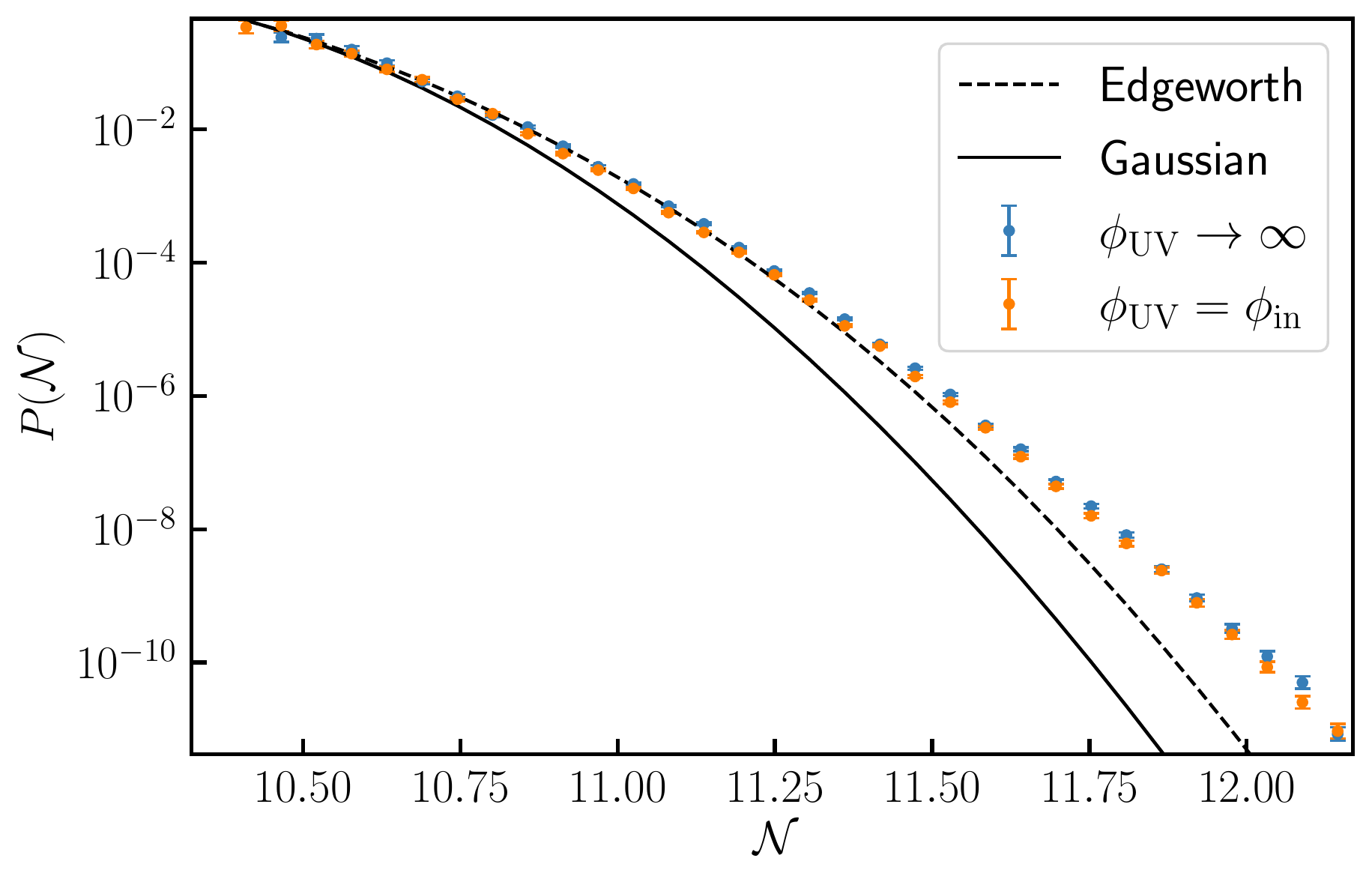}
        \caption{\pdfopening for two locations of the reflective boundary, $\phi_{\mathrm{UV}} = \phi_{\uin} = \sqrt{42}\Mp$ and $\phi_{\mathrm{UV}} \rightarrow \infty$, in quadratic inflation (\ref{eq_quadratic_inflation_potential}) with interim mass $m=0.1\, \Mp$ and bias amplitude $\A = 1.5$. The Gaussian and Edgeworth approximations (see \App{sec_app_edgeworth}) are displayed with solid and dashed black curves, respectively.}
        \label{fig_medium_mass_not_bc_dependent}
\end{center}
\end{figure}

The interim  mass case is phenomenologically interesting, as it produces a PDF, $P(\N)$, with a Gaussian peak and non-Gaussian tail. This is therefore a regime where stochastic effects might be expected to predict an increased PBH production without disturbing the predictions of linear perturbation theory. While one can use the third and fourth central moments of $\N$ in sub-dominant diffusion regime, calculated using the saddle-point approximation~\cite{Vennin:2015hra}, to describe corrections to the Gaussian distribution in the near-tail behaviour using the Edgeworth expansion, see Appendix \ref{sec_app_edgeworth} for details, the far tail has no analytical prediction. This is because central moments beyond the fourth are increasingly difficult to calculate. Therefore, $m \lesssim \Mp$ provides a useful test case for \code, as we can see if it reproduces the analytic predictions at the peak of the PDF and for the near-tail before deviating further into the tail. 

The far tail can be probed with importance sampling (even in the limit $\phi_{\mathrm{UV}} \rightarrow \infty$), as diffusion is sub-dominant and the lognormal estimator for the PDF,  \eq{eq_lognormal_bin_pdf_reconstrcuted_weighted_maximum_likelihood}, can be used. We verified that the $p$-values for the distributions of the logarithms of the weights, $\ln (w)$, in each bin are above the $0.5\%$ threshold.

In \fig{fig_medium_mass_not_bc_dependent} we show the PDF of \fpt when $m=0.1\,\Mp$ and with a bias amplitude $\A = 1.5$, which allows us to probe down to $P(\N)\sim 10^{-10}$ using only $2 \times 10^5$ runs. The PDF is displayed for the two most extreme locations of the reflective boundary, $\phi_{\mathrm{UV}} = \phi_{\uin}$ and $\phi_{\mathrm{UV}} \rightarrow \infty$ (in the later case, in practice, we set $\phi_{\mathrm{UV}} = 100\,\phi_{\uin}$ and find no simulations ever reach this high boundary). Any other choice for $\phi_{\mathrm{UV}}$ leads to results that lie between these two extremes. While there appears to be a slight shift to larger $P(\N)$ at large $\N$ for $\phi_{\mathrm{UV}}\rightarrow \infty$ compared to $\phi_{\mathrm{UV}} = \phi_{\uin}$, the data sets are broadly similar, suggesting that the dependence on $\phi_{\mathrm{UV}}$ appears at even larger $\N$ [hence smaller $P(\N)$] for $m \lesssim \Mp$. One may also see that the logarithmic slope is not constant even at large $\N$, which means that several poles still contribute to the PDF, and the asymptotic exponential tail has not yet been reached. We finally note that both simulations agree within the estimated error with the Gaussian and Edgeworth approximations until $\N\sim 10.6$ and $\N\sim 11.25$ respectively, as expected.

\begin{figure}
\begin{center}
        \includegraphics[width=\figurewidth\textwidth]{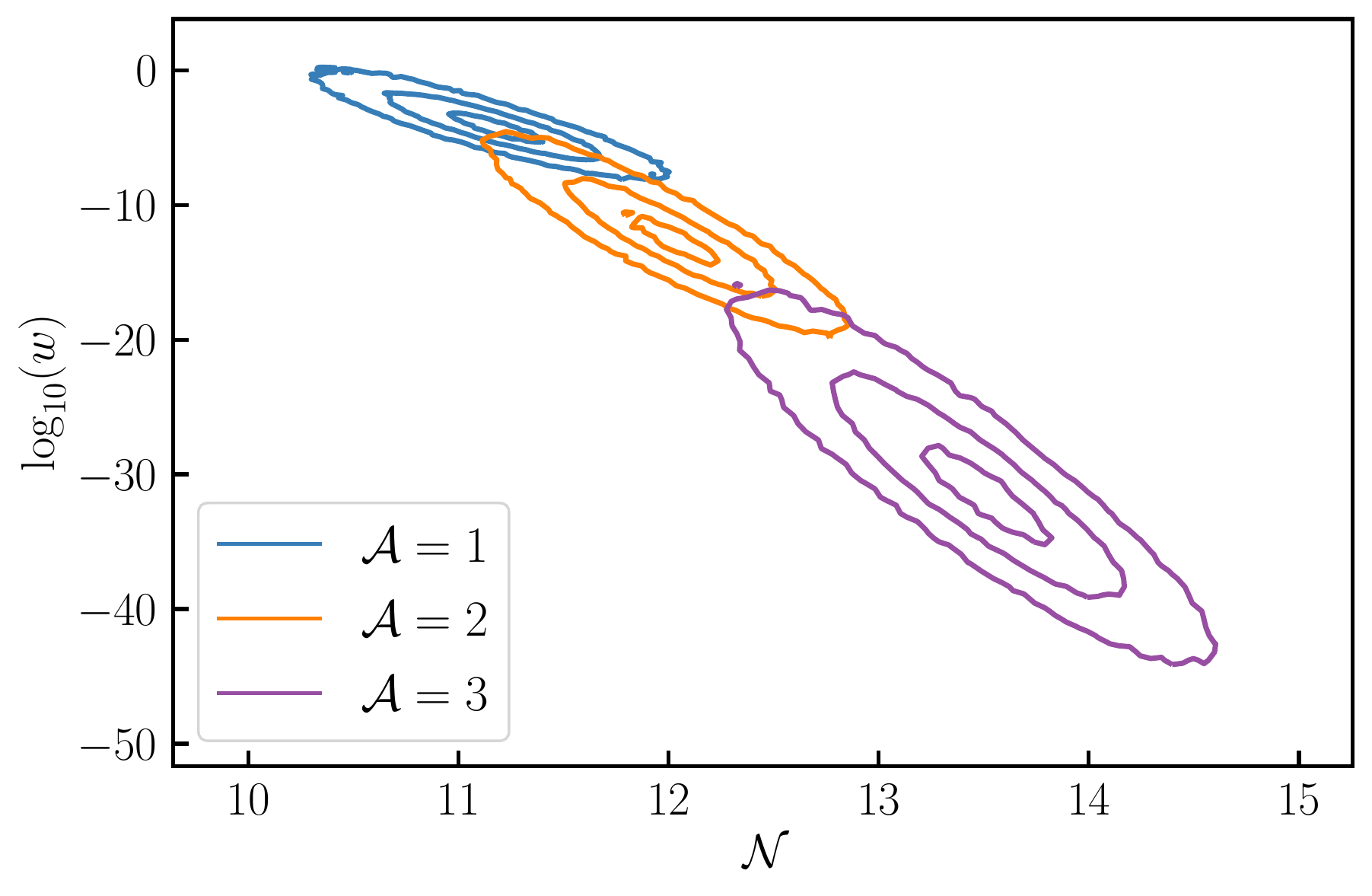}
        \caption{2D histograms for the number of simulation runs in each ${\log}_{10} (w)$ and $\N$ bin represented as contours for quadratic inflation (\ref{eq_quadratic_inflation_potential}) with $m=0.1 \mpl$, $\phi_{\uin}= \sqrt{42}\Mp$ and bias amplitudes $\A=1$ (blue), $\A=2$ (orange) and $\A=3$ (purple). The UV boundary used is $\phi_{\mathrm{UV}} \rightarrow \infty$. Going outwards, the contours represent where 10, 50 and 90 percent of the runs lie, respectively.}
        \label{fig_medium_mass_justifying_combined_plot}
\end{center}
\end{figure}

In \Fig{fig_medium_mass_justifying_combined_plot}, the dependence of the result on the bias amplitude $\A$ is explored, in terms of the distributions of $\N$ and of the weights $w$. When increasing $\A$, one can see that the size of the contours only mildly grow, the main effect being that the location on which they are centred shifts to larger $\N$ and smaller $w$. This confirms that, as discussed around \Eq{eq:meanN:S:gen}, different bias amplitudes allow one to probe different regions of the PDF, since the sample process peaks around different values of $\N$.  This suggests a simple method for reconstructing $P(\N)$ over a wide range, where simulations with different bias amplitudes are combined. This contrasts with the diffusion-domination case where we found that the contours become substantially wider when increasing $\A$, see \Fig{fig_large_mass_case_finite_phi_UV_scatter}. Therefore, when increasing the bias, one indeed  can probe further into the tail even as the number of runs is fixed, but at the expense of mildly diluting the data points into a wider region of $\N$.

\begin{figure}
\begin{center}
        \includegraphics[width=\figurewidth\textwidth]{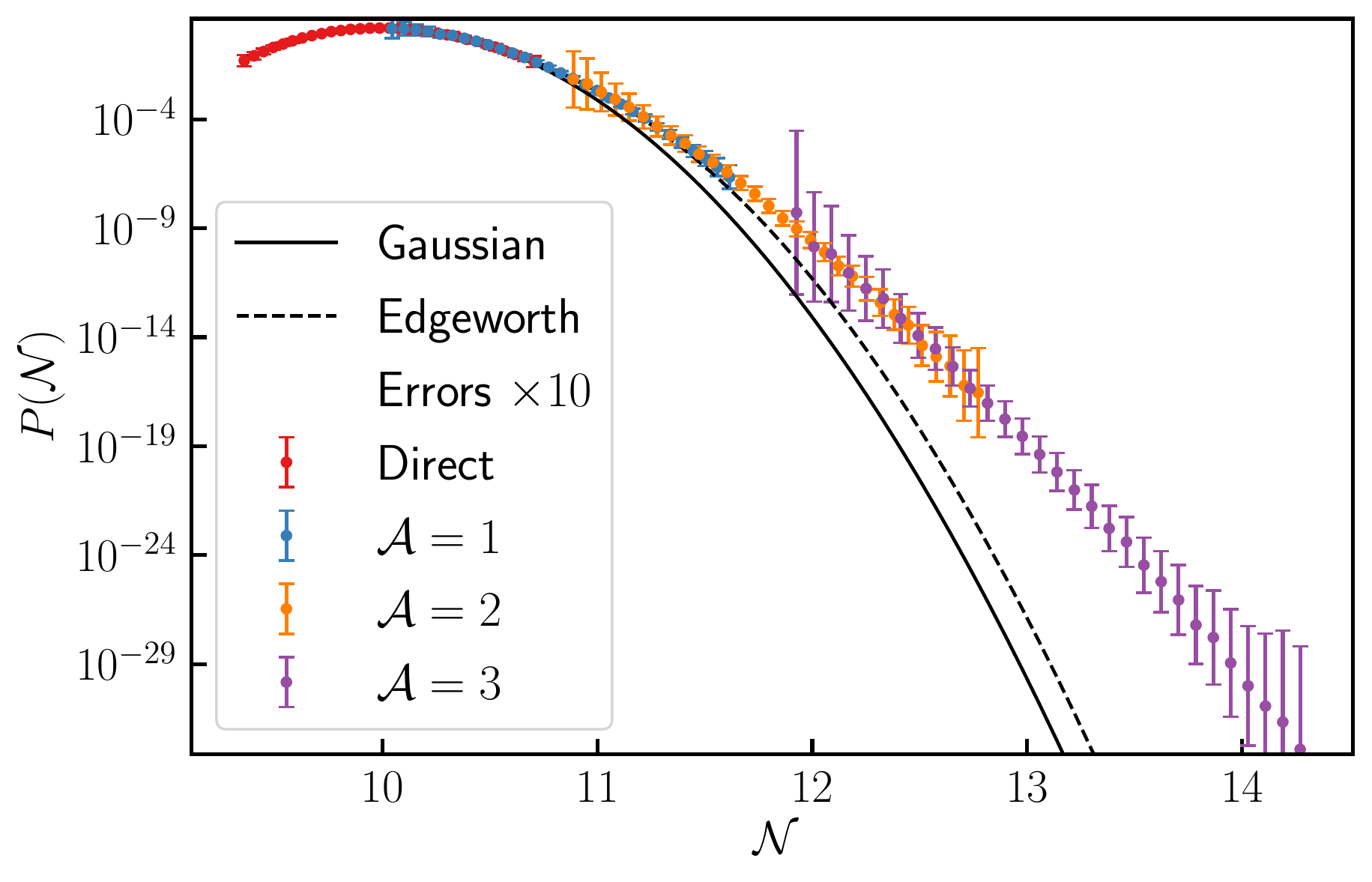}
        \caption{\pdfopening for the data shown in \fig{fig_medium_mass_justifying_combined_plot}, along with $\A=0$ (red, direct sampling). To improve their visibility, {\em the error bars have been magnified by a factor of 10}, \ie their length corresponds to 10 times the error in $\mathrm{log}_{10}[P(\N)]$. The Gaussian and Edgeworth approximations (see \App{sec_app_edgeworth}) are displayed as solid and dashed black curves, respectively.
        }
        \label{fig_medium_mass_combined}
\end{center}
\end{figure}

The multi-bias reconstruction method is employed in \Fig{fig_medium_mass_combined}, where the PDF of \fpt is displayed for $m=0.1\,\Mp$ and several bias amplitudes $\A$. We have magnified the error bars by a factor of 10, otherwise they would not be visible by eye, which shows how efficient the reconstruction is. Where they overlap, we see that the simulations are consistent with one another within the estimated errors and they are in agreement with the Edgeworth expansion for small $\N$, before the expected deviation in the far tail. This clearly demonstrates the efficiency of the importance sampling method, as we required only $10^6$ runs in total to investigate all the way down to $P(\N) \sim 10^{-33}$,  when a direct sample would have required at least $10^{29}$ simulations to reach this far into the tail. A direct sample with $2\times10^5$ runs was only able to investigate down to $P(\N)\sim 10^{-4}$ and $\N=10.66$, which does not reach the point where there is a significant deviation from Gaussianity at $\N=10.91$. Note that as in \Fig{fig_medium_mass_not_bc_dependent}, the exponential tail is still not reached in \fig{fig_medium_mass_combined} (\ie the logarithmic slope is not yet constant at large $\N$), although the tail is explored down to $P(\N) \sim 10^{-33}$ and strong, non-perturbative deviations from Gaussianity are observed. This was also the case when $\phi_{\mathrm{UV}} = \phi_{\uin}$ was investigated.

Let us also note that the size of the error bars still increases with $\A$ in \Fig{fig_medium_mass_combined}, and the reason is twofold. First, the mild increase in the dispersion of the weights noticed in \Fig{fig_medium_mass_justifying_combined_plot} leads to larger error, since $\hat{\epsilon}_j^\pm$, grows with $\hat{\sigma}_{\ln w_j}$ in \Eq{eq_log_normal_bin_pdf_reconstrcuted_error_maximum_likelihood}. Second, when the bias increases the dispersion in $\N$ also increases, which means that fewer data points are found in each bin, and this also results in a larger error since $\hat{\epsilon}_j^\pm$ decreases with $n_j$ (and more precisely as $1/\sqrt{n_j}$ if $n_j$ is large) in \Eq{eq_log_normal_bin_pdf_reconstrcuted_error_maximum_likelihood}. However, let us stress that the two effects have different implications for the quality of the reconstruction: spreading the weight leads to increased error bars, while spreading $\N$ simply leads to diluting the information across a wider region of the PDF.

\subsection{Drift domination: $m \ll \Mp$}
\label{sub_drift_dom}
Let us now turn to the drift-dominated regime where $m \ll \Mp$. We find that a Gaussian distribution remains an excellent approximation for $P(\N)$ up to 15 standard deviations away from the mean (for $m=0.001\, \Mp$ and $\Nmean=10$). As diffusion is even more sub-dominant than in the interim-mass case discussed above, the D'Agostino and Pearson's test is always passed and the lognormal estimator for $P(\N)$ is employed. The multi-bias reconstruction method can also be used. As the interim-mass regime gave results with little dependence on the choice of $\phi_{\mathrm{UV}}$, we expect that the drift-dominated case will be approximately independent of the UV cutoff (which we have verified in practice) and therefore only the results obtained for $\phi_{\mathrm{UV}}\rightarrow \infty$ are shown. 

\begin{figure}
\begin{center}
        \includegraphics[width=\figurewidth\textwidth]{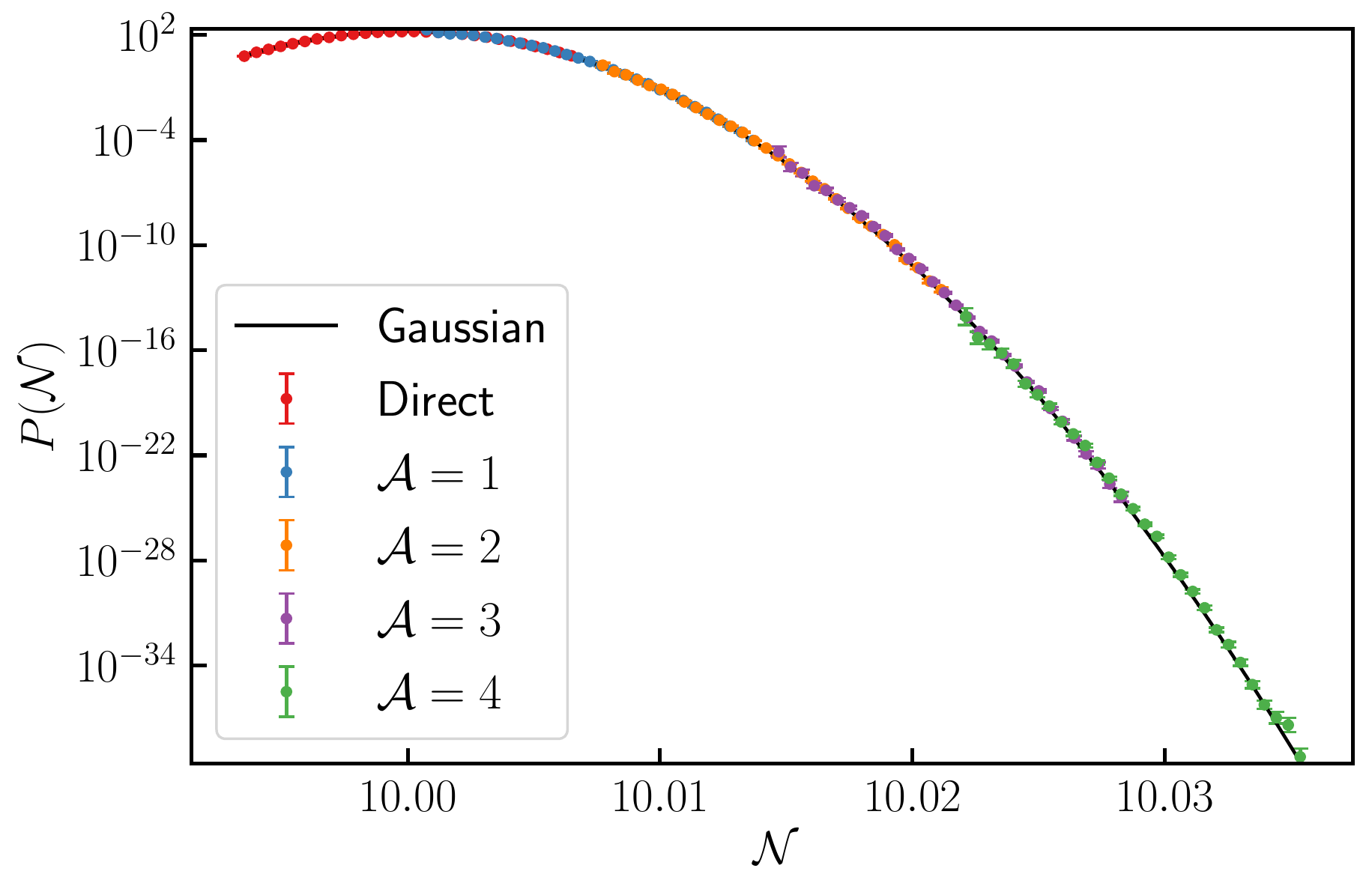}
        
        \caption{\pdfopening for $\phi_{\mathrm{UV}} = \phi_{\uin}$, in quadratic inflation (\ref{eq_quadratic_inflation_potential}) with $m=0.001 \Mp$, $\phi_{\uin}= \sqrt{42}\Mp$ and bias amplitude $\A=0$ (red, direct sampling), $\A=1$ (blue), $\A=2$ (orange), $\A=3$ (purple) and $\A=4$ (green). The Gaussian approximations (see \App{sec_app_edgeworth}) is displayed with a \gaussiancurve~curve.}
        \label{fig_small_mass_combined}
\end{center}
\end{figure}

In \fig{fig_small_mass_combined} we show the results for a series of simulations with varying bias amplitude $\A$. The simulations are consistent with one another when they overlap and the Gaussian approximation within the estimated error. The \code package is able to accurately reproduce $P(\N)$ from the peak of the distribution down to $P(\N) \sim 10^{-39}$ at $13.8$ standard deviations away from the mean, using only $10^6$ runs (while direct sampling would require at least $10^{39}$ runs, and would take more than the age of the universe to complete). This demonstrates the accuracy of \code and suggests that it can be used for exploring the very far tail of the distribution in the drift-dominated regime.

\subsection{Bias optimization}
\label{sec:BiasOptimisation}

\begin{figure}
\begin{center}
        \includegraphics[width=\halffigurewidth\textwidth]{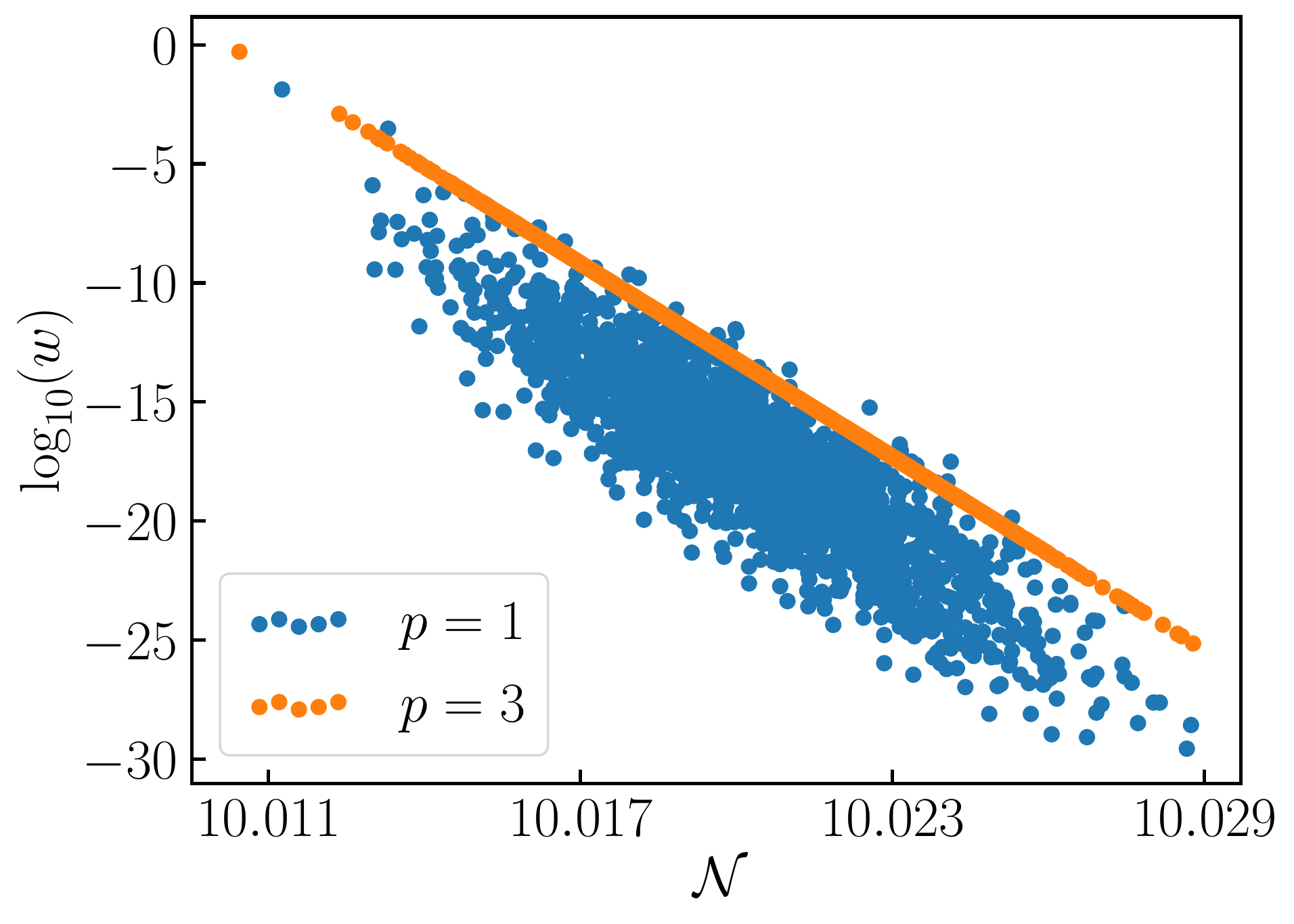}
        \includegraphics[width=\halffigurewidth\textwidth]{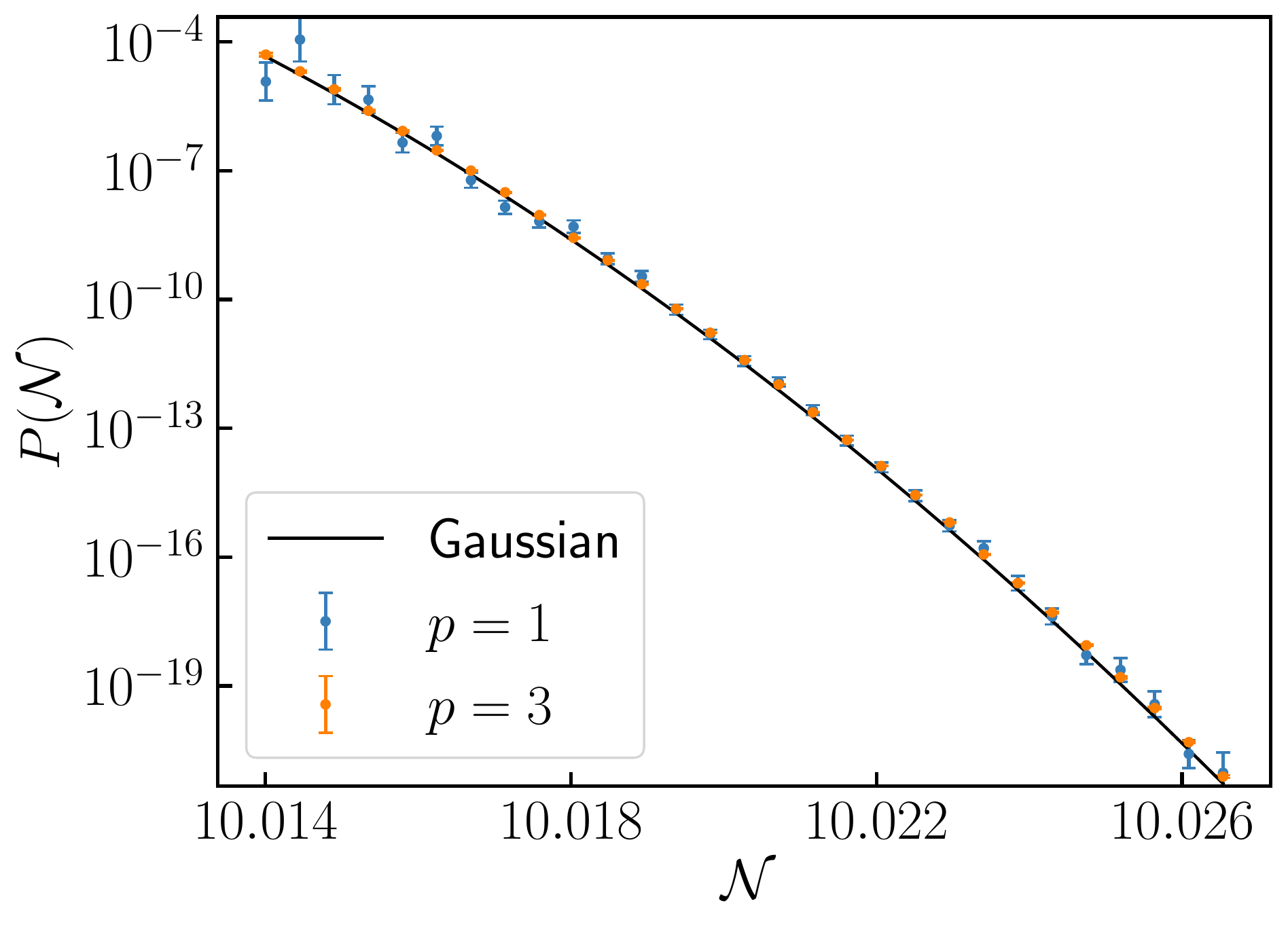}
        \caption{Left panel: scatter plot in ${\log}_{10} (w)$ and $\N$ for quadratic inflation (\ref{eq_quadratic_inflation_potential}) with $m=0.001 \Mp$ and different powers of $p$ in the bias function~\eqref{eq_importance_sampling_amplitude_power_law}. Only $2\times 10^3$ runs are shown for clarity. The value of $\A_p$ is tuned such that the mean of the sample distribution, $\langle \N \rangle_{\mathrm S}$, is 8 standard deviations away from the mean of the target distribution. Right panel: comparison of the first-passage time PDFs, $P(\N)$, for the two data sets. The $p=1$ set uses the lognormal estimator, see \Sec{sub_lognormal_estimator}, while the $p=3$ set is analysed with the na\"ive estimator, see \Sec{sub_importance_sampling}. Both simulation sets use $4 \times 10^4$ runs.}
        \label{fig_small_mass_bias_optimisation_results}
\end{center}
\end{figure}

So far we have used the bias function $f(\phi)=S(\phi)=H(\phi)/(2\pi)$, and have considered how the region of the PDF being sampled can be selected by hand tuning the bias amplitude $\A$. However, in the drift-dominated case, the reconstruction is still numerically expensive. This is because, when solving the Langevin equation, one needs to work with a time step $\Delta N$ that is smaller than the standard deviation of $\N$, and that standard deviation becomes tiny when quantum diffusion is subdominant (in the present model it grows linearly with $m$). Therefore many steps are required. This prompts us to explore whether further efficiency can be gained by optimising the bias function, $f(\phi)$, for the inflation model under investigation. This is why in this section we consider a more general, power-law form of the bias function
\begin{equation}
   \label{eq_importance_sampling_amplitude_power_law}
    \B(\phi) = \A_p \phi^p \,.
\end{equation}
The previous bias function for quadratic inflation (\ref{eq_quadratic_inflation_potential}) corresponds to the choice $p=1$, since from \eq{eq_hubble_sr}, $H(\phi)\propto \phi$.

In \Fig{fig_small_mass_bias_optimisation_results}, we compare the previous bias function ($p=1$) with the case where $p=3$, which we will see below is the optimal choice. The left panel displays the scatter in $\N$ and $w$ for $2 \times 10^3$ data points, and one can see that the choice $p=3$ leads to a distribution that is substantially more squeezed, hence values that are more correlated. This implies that, within a given $\N$ bin, the dispersion of the weights is smaller. This leads to reduced errors, as confirmed in the right panel of \Fig{fig_small_mass_bias_optimisation_results} where the PDFs corresponding to these two values of $p$ are displayed. Note that the typical values of $\ln(w)$ are smaller for $p=1$ than for $p=3$ in the left panel. This is because, to (eventually) converge on the same $P(\N)$ in each bin, \Eq{eq_lognormal_mean_maximum_likelihood} indicates that $\langle \ln w\rangle$ must be smaller if $\sigma_{\ln w}$ is larger.

The dispersion of the weights is so small in the $p=3$ case that the variance in $\ln w$, $\sigma_{\ln w}$, is difficult to resolve. This means that the lognormal estimate for the error, \Eq{eq_log_normal_bin_pdf_reconstrcuted_error_maximum_likelihood}, is difficult to evaluate, which explains why we rather use the na\"ive estimate for $p=3$. The two estimates anyway converge for vanishingly small $\sigma_{\ln w}$. The right panel of \fig{fig_small_mass_bias_optimisation_results} was produced using only $4 \times 10^4$ runs in each simulation. For $p=3$ we obtain results that are as good as the lognormal data points shown in the right panel of \Fig{fig_showing_scatter_in_w_and_error}, for a simulation with $2 \times 10^5$ runs. For $p=1$, larger error bars are obtained.

\begin{figure}
\begin{center}
        \includegraphics[width=\figurewidth\textwidth]{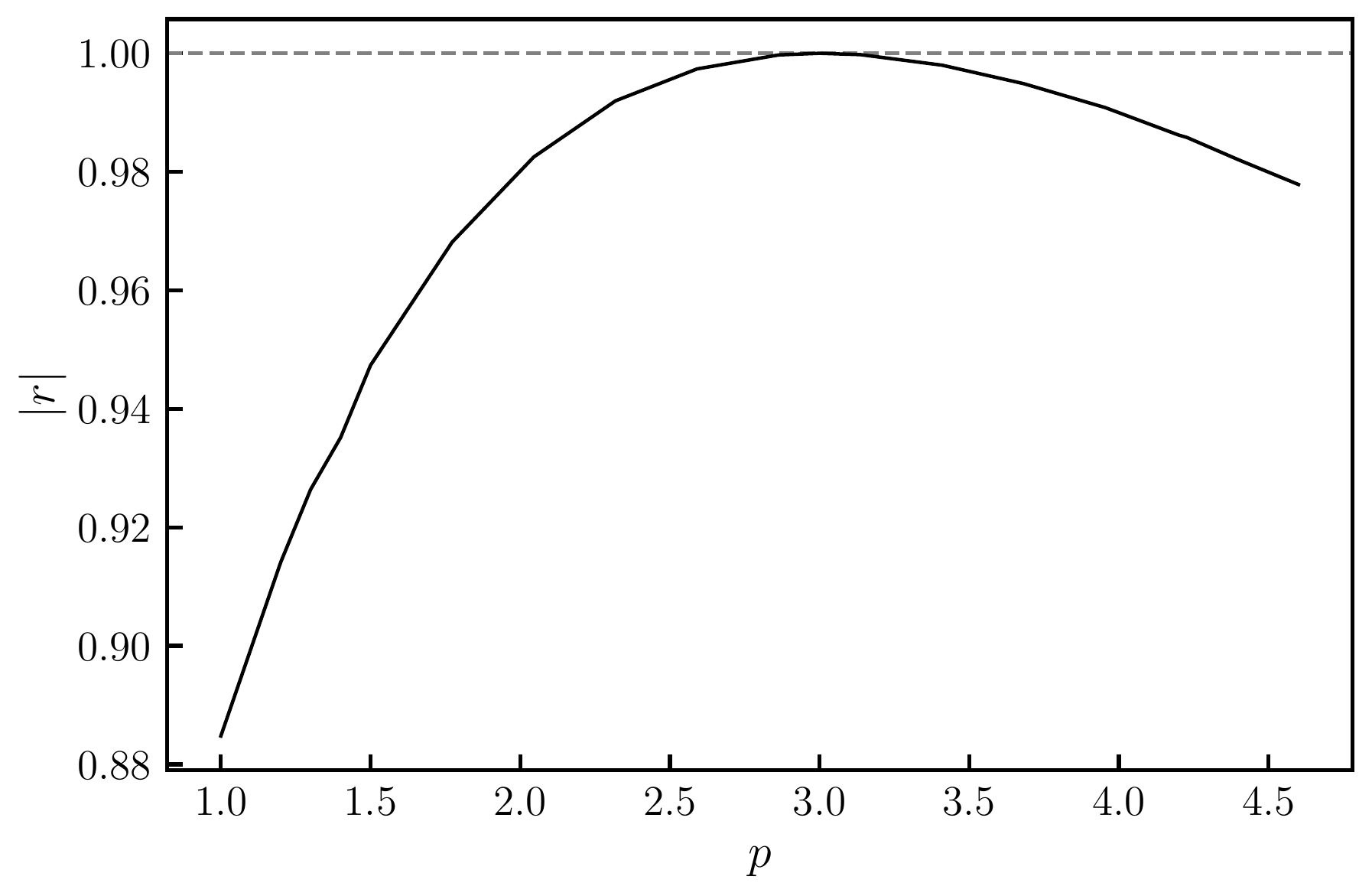}
        
        \caption{The modulus of Pearson's correlation coefficient~\cite{freedman1998statistics}, $|r|$, of $\log_{10} (w)$ with $\N$ for different values of the power $p$ appearing in the bias function given in \eq{eq_importance_sampling_amplitude_power_law}, for quadratic inflation (\ref{eq_quadratic_inflation_potential}) with $m=0.001\, \Mp$. The dashed gray line shows $|r|=1$, corresponding to perfect correlation between  $\log_{10} (w)$ and $\N$.} 
        \label{fig_small_mass_bias_optimisation_r_values}
\end{center}
\end{figure}

This suggests a method for optimising the value of $p$, by measuring how much the $(\N,w)$ contours are squeezed. This can be done by computing Pearson's correlation coefficient~\cite{freedman1998statistics} between $\log_{10} (w)$ and $\N$, denoted $r$, which vanishes for uncorrelated variables and equals $1$ and $-1$ for fully correlated and anti-correlated variables respectively. The result is displayed in \Fig{fig_small_mass_bias_optimisation_r_values}, and confirms that the optimal value is indeed around $p\simeq 3$.

When the bias function is optimised, the different estimators (naive and lognormal) give similar results and can be used equivalently. However, the bias-optimisation programme may not be tractable beyond simple models like the one investigated here, so in general it remains important to choose the estimator appropriately.

\section{Conclusions}
\label{sec_conclusions}

In this work, we have shown how importance sampling can be used to investigate rare
cosmological fluctuations in stochastic inflation. We have developed the publicly available \code package, which reconstructs the far tail of the probability distribution of first-passage times for general one-dimensional Langevin processes. Importance sampling is achieved by introducing a bias into the simulated Langevin equation, to increase the occurrence of rare realisations of the stochastic process. As by the stochastic-$\delta N$ formalism the local duration of inflation in \efolds, $\N$, is directly related to the curvature perturbation by $\zeta = \N - \Nmean$, finding the PDF of $\zeta$ (and hence of $\N$) corresponds to solving a first-passage time problem.

Testing the robustness of importance-sampling methods for stochastic inflation was our main objective. To this end, quadratic inflation was used as a benchmark model that provides analytical or semi-analytical comparisons to the numerical results. This confirmed the accuracy of the \code package within the estimated errors. More precisely, we found that for diffusion domination (obtained within quadratic inflation by setting $m \gtrsim \Mp$), with $2 \times 10^5$ simulation runs, importance sampling allows one to reconstruct the PDF of $\N$ down to $P(\N) \sim 10^{-12}$, while direct sampling is only able to reach $P(\N) \sim 10^{-3}$, see \fig{fig_large_mass_case_finite_phi_UV}. In the interim case where $m \lesssim \Mp$, and in the drift-domination regime ($m \ll \Mp$), importance sampling allows the PDF down to $P(\N) \lesssim 10^{-33}$ to be simulated, see \Figs{fig_medium_mass_combined} and \ref{fig_small_mass_combined} respectively. Randomly realising just one of these events without importance sampling would take longer than the age of the universe with current super computers. Here, only a laptop CPU was used.

The dependence of the statistics of rare fluctuations on the UV reflective boundary, $\phi_{\mathrm{UV}}$, was also investigated. In the diffusion-dominated regime, the first-passage time distribution carries a significant dependence on $\phi_{\mathrm{UV}}$, although this dependence is removed to further into the tail as $\phi_{\mathrm{UV}}$ increases, see \Fig{fig_large_mass_case_phi_UV_comparisons}. The asymptotic exponential tail of the distribution, \ie the part  dominated by the leading pole of the characteristic function, is also found deeper into the tail as $\phi_{\mathrm{UV}}$ increases, and the seemingly exponential tails in \Fig{fig_large_mass_case_phi_UV_comparisons} are in fact given by the next-to-leading pole. In the interim and drift-dominated regimes, no dependence on $\phi_{\mathrm{UV}}$ is observed in the range probed [namely down to $P(\N) \sim 10^{-33}$, see \Fig{fig_medium_mass_not_bc_dependent}]. In the interim case, we still observe non-perturbative deviations from Gaussianity in the tail, although they are not of the simple exponential type, denoting simultaneous contributions from multiple poles. This also indicates that, in regimes where quantum diffusion does not dominate, it may not be enough to approximate the non-Gaussian tail with an exponential fit. In the drift-dominated regime, we find no significant deviation from Gaussianity down to probabilities $P(\N) \sim 10^{-39}$, see \Fig{fig_small_mass_combined}. 

By tuning the bias, one can select which part of the distribution is to be best reconstructed. In the interim and drift-dominated cases, this can be used to probe almost arbitrarily far in the tail\footnote{We could have investigated as far as $P(\N) \sim 10^{-65}$, but beyond this point there would be less than one such Hubble patch in the entire observable universe.}. This is because the distributions of the weights, the relative probability of realising importance sample without the bias, were found to be given by lognormal distributions. This allows one to design estimators of the PDF that benefit from all realisations within a given bin, even those with small weight.

We also found that the functional form of the bias could be optimised to the potential investigated and have solved this optimisation problem for quadratic inflation, where the variance in the weights within each $\N$ bin was minimized (see \fig{fig_small_mass_bias_optimisation_results}). This suggests that for numerically expensive cases, such as drift domination, further efficiency can be gained by choosing a suitable form for the bias. This also has the additional advantage of making lognormal estimator unnecessary, as the na\"ive method is sufficient for vanishing weight variance. Of course this requires an optimized form for the bias, which is not in general known, while the lognormal estimator can be employed for all drift dominated cases. Finding the optimal bias function in a generic setup is an open issue, which would potentially allow us to increase the efficiency of importance sampling, and requires further investigation.

In the diffusion-dominated case, the distribution of the weights is more involved and cannot be modeled with a lognormal law. This implies that a more ``na\"ive'' estimator must be used, which is dominated by the few runs with largest weight. This leads to larger statistical error, so in practice one is limited in how far into the tail one can go.

The next step is to extend \code to solve multi-dimensional first-passage time problems. This would allow investigations beyond the single-field slow-roll approximation to be done, such as phases of ultra-slow-roll~\cite{Firouzjahi:2018vet, Ballesteros:2020sre, Pattison:2021oen} or multiple-field setups~\cite{Garcia-Bellido:1996mdl, Martin:2011ib, Clesse:2015wea, Kawasaki:2015ppx, Assadullahi:2016gkk, Vennin:2016wnk, Noorbala:2019kdd, Pinol:2020cdp, Hooshangi:2022lao}. In principle, the noise amplitude is determined by the field mode equation when integrated along a given realisation, and this leads to non-Markovian effects~\cite{PerreaultLevasseur:2013eno, Figueroa:2021zah} that we could also incorporate, together with other recently-proposed refinements of the stochastic-inflation equations~\cite{Cohen:2021fzf,Mahbub:2022osb}.

Finally we note that to make direct contact with observational constraints such as the abundance of primordial black holes, one still needs to map the distribution of first-crossing times at the end of inflation to the compaction function at a given scale, when primordial curvature perturbations are highly non Gaussian~\cite{Ando:2020fjm, Tada:2021zzj,Biagetti:2021eep, Kitajima:2021fpq}. This important step is left for future work.

\section*{Data Availability Statement} 

The data from which the results presented here are derived is freely available at \url{https://github.com/Jacks0nJ/PyFPT}, along with the code required to produce the data.
\acknowledgments

We would like to thank Andrew Gow, Chris Pattison, Lucas Pinol, Sébastien Renaux-Petel, Lukas Witkowski and the participants of the Paris-Portsmouth mini workshop for insightful discussions. JJ would like to thank Ian Harry and Coleman Krawczyk for help in developing the code into a package, as well as APC Paris for their hospitality during his Long Term Attachment. This work was supported by the Science and Technology Facilities Council (grant numbers ST/T506345/1, ST/S000550/1 and ST/W001225/1).

For the purpose of open access, the authors have applied a Creative Commons Attribution (CC-BY) licence to any Author Accepted Manuscript version arising from this work.

\appendix

\section{Semi-analytical results}
\label{sec_app_analysis}

To test the importance-sampling method, it is useful to compare its result with analytical or semi-analytical predictions. In this appendix, we derive such predictions for the simple model studied in this work, namely quadratic inflation~\eqref{eq_quadratic_inflation_potential}.


Following the approach laid out in \Refa{Pattison:2017mbe}, we first
introduce the characteristic function
\begin{equation}
    \label{eq_appendix_chracteristic_function}
    \chi_{\N}(t, \phi) = 
    \int_{-\infty}^{\infty} e^{it\N } P_{\phi}(\N) \d  \N \, ,
\end{equation}
where $t$ is a dummy variable. In stochastic slow-roll inflation, the characteristic function obeys the adjoint Fokker-Planck equation~\cite{Pattison:2017mbe}
\begin{equation}
    \label{eq_chracteristic_function_partial_diff_eqn}
    \bigg(  \frac{\partial^2}{\partial \phi^2} - \frac{v'}{v^2}  \frac{\partial}{\partial \phi} + \frac{it}{v\Mp^2}  \bigg) \chi_{\N}(t, \phi) = 0 \, ,
\end{equation}
where the reduced potential 
\begin{equation}
    \label{eq_reduced_potential}
    v(\phi) = \frac{V(\phi)}{24 \pi^2 \Mp^4} 
\end{equation}
was introduced to simplify notations. Due to the absorbing boundary at $\phi_{\mathrm{end}}$ and the reflective boundary at $\phi_{\mathrm{UV}}$, the FPT distribution must be such that $P_{\phi_\uend}(\mathcal{N})=\delta(\mathcal{N})$ and $\frac{\partial}{\partial\phi} P_{\phi_{\mathrm{UV}}}(\mathcal{N})=0$, hence the characteristic function must satisfy the boundary conditions
\begin{equation}
\label{eq_chracteristic_function_bc}
    \chi_{\N}(t,\phi_{\mathrm{end}}) = 1 \quad \text{and} \quad \frac{\partial \chi_{\N}}{\partial \phi}(t,\phi_{\mathrm{UV}}) = 0 \, .
\end{equation}
In general, the solution to \Eq{eq_chracteristic_function_partial_diff_eqn} under the conditions~\eqref{eq_chracteristic_function_bc} can be cast in the form~\cite{Ezquiaga:2019ftu}
\begin{equation}
    \label{eq_chracteristic_function_expanded}
    \chi_{\N}(t,\phi) = \sum_n \frac{a_n(\phi)}{\Lambda_n - it} + g(t, \phi) \, , 
\end{equation}
where $a_n(\phi)$ and $\Lambda_n$ are the $n^{\mathrm{th}}$ residual and pole respectively, and $ g(t, \phi)$ is a regular function. By inverse Fourier transforming \Eq{eq_appendix_chracteristic_function}, one obtains
\begin{equation}
\label{eq:PDF:from:chi}
P_\phi(\N) = \frac{1}{2\pi}\int_{-\infty}^\infty \ee^{-it\N}\chi_{\N}(t,\phi)\dd t\, .    
\end{equation}
Computing the integral with the residue theorem, one then finds
\begin{equation}
     \label{eq_pdf_expansion_in_exponentials2}
     P_{\phi}(\N) = \sum_n a_n(\phi)e^{-\Lambda_n \N} \, ,
\end{equation}
\ie $P_\phi(\N)$ is a sum of decaying exponentials.


Let us now apply this program to quadratic inflation,
\bea
\label{eq_appendix_rescaled_potential}
v=v_0\left(\frac{\phi}{\Mp}\right)^2 \, ,
\eea
see \Eq{eq_quadratic_inflation_potential}, where inflation takes place between $\phi_\uend=\sqrt{2}\Mp$ (where inflation ends by slow-roll violation, hence an absorbing wall is set) and $\phi_\uUV$ (where a reflective wall is set). Substituting the potential function~\eqref{eq_appendix_rescaled_potential} into \Eq{eq_chracteristic_function_partial_diff_eqn}, one obtains
\bea
\left(\frac{\partial^2}{\partial\phi^2}-\frac{2\Mp^2}{v_0\phi^3}\frac{\partial}{\partial\phi}+\frac{i t}{v_0 \phi^2}\right)\chi\left(t,\phi\right)=0\, ,
\eea
the generic solution of which is given by
\bea
\chi\left(t,\phi\right) = \lambda(t) f\left(t,\phi\right) + \mu(t) g\left(t,\phi\right),
\eea
where
\bea
\label{eq_appendix_f_and_g}
f\left(t,\phi\right) &= \left[v(\phi)\right]^{\frac{1+\alpha(t)}{4}} \, {}_1F_1 \left[\frac{-1-\alpha(t)}{4};1-\frac{\alpha(t)}{2};-\frac{1}{v(\phi)}\right] \, , \\
g\left(t,\phi\right) &= \left[v(\phi)\right]^{\frac{1-\alpha(t)}{4}} \, {}_1F_1 \left[\frac{-1+\alpha(t)}{4};1+\frac{\alpha(t)}{2};-\frac{1}{v(\phi)}\right] \, .
\eea
Here, $\lambda(t)$ and $\mu(t)$ are two integration constants, and to simplify notations we have introduced the parameter
\bea
\label{eq_appendix_alpha}
\alpha(t)=\sqrt{1-4\frac{it}{v_0}} \, .
\eea
The boundary conditions given in \eq{eq_chracteristic_function_bc} imply that
\bea
\label{eq_mu_and_lambda}
\mu  = \frac{B}{B-A}
\quad\text{and}\quad
\lambda &= -\frac{A}{B-A}\, ,
\eea
where
\bea
\label{equation_appendix_A_and_B}
A\equiv \frac{f_\uUV'}{g_\uUV'} \quad\text{and}\quad
B\equiv \frac{f_\uend}{g_\uend}\, .
\eea
Here, $f_\uend$ is a shorthand for $f(t,\phi_\uend)$ etc, and a prime denotes a derivation with respect to the field value. The characteristic function can thus be written as
\bea
\label{eq:characteristicFunction:generic}
\chi\left(t,\phi\right) =
\frac{B(t)}{B(t)-A(t)} \frac{f(t,\phi)}{f_\uend(t)}-\frac{A(t)}{B(t)-A(t)}\frac{g(t,\phi)}{g_\uend(t)} \, .
\eea

\subsection*{Numerical reconstruction}
\label{sub_app_numerical_reconstruction}

Two methods can then be used to reconstruct the FPT distribution function. The first one consists in performing the integral in \Eq{eq:PDF:from:chi} numerically.
The second method relies on numerically solving the pole equation
\begin{equation}
    \label{eq_appendix_pole_equation}
    A(t) = B(t)
\end{equation}
and derive the corresponding values of the poles $\Lambda_n=it_n$. 
The residues can then be obtained with the formula
\begin{equation}
    \label{eq_appendix_residual_numerical}
    a_n(\phi) = \lim_{t \to -i \Lambda_n} (\Lambda_n - i t) \frac{B(t)\frac{f(t, \phi)}{f_{\mathrm{end}}(t)}- A(t)\frac{g(t, \phi)}{g_{\mathrm{end}}(t)}}{B(t)-A(t)} \, ,
\end{equation}
which can be further simplified using L'H\^{o}pital's rule, leading to
\begin{equation}
    \label{eq_appendix_residual_simplified}
    a_n(\phi) =   \lim_{t \to -i \Lambda_n} -i  \frac{ B(t)\frac{f(t, \phi)}{f_{\mathrm{end}}(t)}- A(t)\frac{g(t, \phi)}{g_{\mathrm{end}}(t)} }{\frac{\partial}{\partial t} \left[B(t)-A(t)\right]}.
\end{equation}
The PDF can then be evaluated with \Eq{eq_pdf_expansion_in_exponentials2}, \ie
\bea 
\label{eq_appendix_pdf_for_residuals}
P_\phi(\N)=\sum_n \lim_{t \to -i \Lambda_n} -i  \frac{ B(t)\frac{f(t, \phi)}{f_{\mathrm{end}}(t)}- A(t)\frac{g(t, \phi)}{g_{\mathrm{end}}(t)} }{\frac{\partial}{\partial t} \left[B(t)-A(t)\right]} \ee^{-\Lambda_n \N}\, .
\eea 

In practice, only a finite number of poles $\Lambda_n$ for $n\in \llbracket 0,n_\umax\rrbracket$ is extracted with their corresponding residues. One then needs to check that $n_\umax$ is large enough for the reconstructed PDF to not depend on it. We find that such a number required for an accurate estimation depends not only on $m$ and $\phi_{\mathrm{UV}}$ but also on how far into the tail one is approximating.

For diffusion domination with $m \gtrsim \Mp$ and a low energy UV cutoff such as $\phi_{\mathrm{UV}} \simeq \phi_\uin$, including the first 50 poles in \eq{eq_appendix_pdf_for_residuals} gives a well-converged approximation even for the peak of the distribution. Further in the tail, only the leading- and next-to-leading poles are needed. However, for the interim case with $m \lesssim \Mp$ and high energy UV cutoff $\phi_{\mathrm{UV}} \gg \phi_\uin$, thousands of poles are required for an accurate estimation even at $P(\N) \sim 10^{-20}$. 
As we shall now see, the reason is the distance between the poles decreases with $\phi_{\mathrm{UV}}$, hence more poles contribute to the result as $\phi_{\mathrm{UV}}$ increases, even deep in the tail.

\subsection*{Large-$\phi_{\mathrm{UV}}$ limit}
\label{sub_app_pole_distance}
The distance between two consecutive poles depends on the value of $\phi_{\mathrm{UV}}$, and to show this property explicitly let us now study the large-$\phi_{\mathrm{UV}}$ regime.  The terms $f'_\uUV$ and $g'_\uUV$ can be approximated by expanding the hypergeometric functions in the limit where their last argument is close to zero while being negative, leading to\footnote{This is obtained by combining Eqs.~(13.3.15), (13.2.39) and (13.2.13) of \Refa{NIST:DLMF}.}
\bea
f'_\uUV &= \left(v_\uUV\right)^{\frac{1+\alpha}{4}} \frac{1+\alpha}{2}\frac{1}{\phi_\uUV}\left[1+\mathcal{O}\left(\frac{1}{v_\uUV}\right)\right],\\
g'_\uUV &= \left(v_\uUV\right)^{\frac{1-\alpha}{4}} \frac{1-\alpha}{2}\frac{1}{\phi_\uUV}\left[1+\mathcal{O}\left(\frac{1}{v_\uUV}\right)\right] ,
\eea
where $v_{\mathrm{UV}}=v(\phi_{\mathrm{UV}})$. Note that for this expansion to be valid, one needs $v_{\mathrm{UV}}$ to be large, which implies $\phi_{\mathrm{UV}}\gg v_0^{-1/2}\sim\Mp^2/m$. For sub-Planckian values of $m$ such field values are very large, and presumably out of the validity range of the model, but here the large-$\phi_{\mathrm{UV}}$ limit is worked out only as a formal way to understand how $\phi_{\mathrm{UV}}$ affects the pole locations.
These expressions lead to
\bea
\label{eq_appendix_A_approx}
A= \frac{f'_\uUV}{g'_\uUV} = \frac{1+\alpha}{1-\alpha} \left(v_\uUV\right)^{\frac{\alpha}{2}}\left[1+\mathcal{O}\left(\frac{1}{v_\uUV}\right)\right]\, .
\eea

To approximate $B=f_\uend/g_\uend$, let us further assume that $v_0\ll 1$, hence $v_\uend\ll 1$ with $\phi_\uend=\sqrt{2}\Mp$. By expanding the hypergeometric functions in the limit where their last argument is very large negative, one obtains\footnote{This is obtained by combining Eqs.~(13.2.39) and (13.2.23) of \Refa{NIST:DLMF}.}
\bea
\label{eq_appendix_B_approx}
B=\frac{f_\uend}{g_\uend} \simeq \dfrac{\Gamma\left(1-\frac{\alpha}{2}\right)\Gamma\left(\frac{5+\alpha}{4}\right)}{\Gamma\left(1+\frac{\alpha}{2}\right)\Gamma\left(\frac{5-\alpha}{4}\right)}\, .
\eea
In this regime the pole equation (\ref{eq_appendix_pole_equation}) reduces to
\bea
\label{eq:PoleCondition:Appr}
{\frac{1+\alpha}{1-\alpha} \left(v_\uUV\right)^{\frac{\alpha}{2}}} = {\dfrac{\Gamma\left(1-\frac{\alpha}{2}\right)\Gamma\left(\frac{5+\alpha}{4}\right)}{\Gamma\left(1+\frac{\alpha}{2}\right)\Gamma\left(\frac{5-\alpha}{4}\right)}}\, .
\eea
This equation needs to be solved on the negative imaginary axis, \ie when $t=-i\Lambda$, where $\Lambda$ is real positive. For $0<\Lambda<v_0/4$, $\alpha$ is real and comprised between $0$ and $1$, see \Eq{eq_appendix_alpha}. In that case, it can then be shown that the pole equation has no solution if $v_\uUV$ is of order one or larger (which we assumed). Moreover, if $\Lambda=v_0/4$, then $\alpha=0$ and the pole equation is trivially satisfied. This thus gives us the leading pole,
\bea
\Lambda_0=\frac{v_0}{4}\, .
\eea
Note that when $\alpha=0$, one has $f=g$ and it is clear that $A=B= 1 $ exactly.
The expression we just derived for the leading pole is therefore valid in general, beyond the large-$\phi_{\mathrm{UV}}$ limit.

Let us now study the location of the first higher poles. When $\Lambda>v_0/4$, $\alpha$ is imaginary, thus both hands of the pole equation~\eqref{eq:PoleCondition:Appr} oscillate. While the left-hand side oscillates in $\alpha$ with a frequency of the order $\ln (v_\uUV)$, the right-hand side oscillates at a slower rate with a frequency of order one. If $\ln(v_\uUV)\gg 1$, the first oscillations in the left-hand side occur while the right-hand side is still approximately constant, so one may replace $\alpha\simeq 0$ in \Eq{eq:PoleCondition:Appr} except in the term $(v_\uUV)^{\alpha/2}$. This leads to
\bea
\label{eq_appendix_approx_pole_equation}
 \left(v_\uUV\right)^{\frac{\alpha}{2}} = 1\, ,
\eea
which, by replacing $t=-i \Lambda$ and substituting in \eq{eq_appendix_alpha}, has solutions
\bea
\label{eq_appendix_approx_pole}
\left. \Lambda_n\right\vert_{n\ll \frac{\ln (v_\uUV)}{4 \pi}} \simeq \frac{v_0}{4} \left[1+\left(\frac{4\pi n}{\ln v_\uUV}\right)^2\right]\, .
\eea
This approximation is valid as long as $\vert\alpha\vert\ll 1$, which implies that $n\ll \ln(v_\uUV)/(4 \pi)$. This is why only the first poles are approached by this formula, which is relevant only when $\ln(v_\uUV)$ is large (hence $\phi_\uUV$ is tremendously large). It however makes explicit that the distance between poles, $\Lambda_{n-1}-\Lambda_n$, decreases with $\phi_\uUV$. This decrease is only logarithmic but it does show that as $\phi_\uUV$ increases, more poles contribute to the PDF. It also means that as $\phi_\uUV$ goes to infinity, the set of poles become continuous, hence the asymptotic tail is not exponential anymore, as we confirm below.
\subsection*{Infinite-$\phi_\uUV$ limit}
\label{sub_app_infinite_phi_UV}
The case where $\phi_\uUV$ is infinite is indeed peculiar and needs to be treated separately. 
When $\phi\to \infty$, $1/v \to 0$ and the hypergeometric functions in \Eq{eq_appendix_f_and_g} tend to a constant. The behaviour of $f$ and $g$ in \eq{eq_appendix_f_and_g} is then dictated by the prefactors $v^{(1\pm \alpha)/4}$. Whether this diverges or not when $v\to\infty$ depends on the sign of $\Rea(1+\alpha)$. We restrict this sign analysis to the case where $t$ is real, recalling that the PDF can be extracted using \Eq{eq:PDF:from:chi}, which only relies on evaluating the characteristic function with real $t$. 

Since $\alpha^2 = 1-4 i t/v_0$, one has $\Rea(\alpha^2)=1$ so can write $\alpha^2 = \ee^{i\theta}/\cos(\theta)$, where $-\pi/2<\theta<\pi/2$. This gives rise to $\alpha=\ee^{i\theta/2}/\sqrt{\cos(\theta)}$, meaning $\Rea(\alpha)= \sqrt{[1+1/\cos(\theta)]/2}$. When $-\pi/2<\theta<\pi/2$, one thus has $\Rea(\alpha)>1$. Therefore, when $v$ goes to infinity,  $f$ diverges while $g$ asymptotes a constant. In order to satisfy the second boundary condition in \Eq{eq_chracteristic_function_bc}, one must thus keep the $g$ branch only and set $\lambda(t)=0$. The remaining integration constant, $\mu(t)$, can then be set in order to satisfy the first boundary condition and this gives rise to
\bea
\label{eq_appendix_characteristic_function_infinite_UV}
\chi\left(t,\phi\right) =\left( \frac{v}{v_\uend}\right)^{\frac{1-\alpha}{4}} \frac{{}_1F_1 \left(\frac{-1+\alpha}{4};1+\frac{\alpha}{2};-\frac{1}{v}\right)}{{}_1F_1 \left(\frac{-1+\alpha}{4};1+\frac{\alpha}{2};-\frac{1}{v_\uend}\right)}\, .
\eea

The PDF can then obtained from \Eq{eq:PDF:from:chi} (a numerical extraction of the poles is indeed impossible, given that they constitute a continuous spectrum when $\phi_\uUV$ is infinite). 
On the negative imaginary axis when $it>v_0/4$, the first derivatives of the characteristic function are discontinuous, \ie there is a branch cut. Together with the fact that there are no poles, we can deform the contour of integration to rewrite \Eq{eq:PDF:from:chi} as \begin{equation}
\label{eq_appendix_pdf_infinite_UV}
    P_{\phi}(\N) = - 2 \int_{v_0/4}^{\infty} 
    \Ima [\chi(-i y, \phi)] e^{- y \N} \dd y.
\end{equation}
This integral can be computed numerically, but requires very high precision for $v_0 \ll 1$ due to the large argument in the confluent hypergeometric function. 

Let us finally note that the far tail of the PDF  at $\N \gg 1/v_0$ is determined by the behaviour of the imaginary part of the characteristic function around $y = v_0/4$. Expanding the characteristic function for small $|\alpha|$, one can show that $\Ima[\chi(-i y, \phi)] \propto \Ima [\alpha(- i y)]$ and $ P_{\phi}(\N) =\N^{-3/2} \exp(- v_0 \N/4)$. Therefore, as announced above, in the case where $\phi_\uUV$ is infinite the tail is not exponential anymore, but only quasi exponential, due to the presence of a continuous spectrum of poles.

\section{Edgeworth expansion}
\label{sec_app_edgeworth}

In the main text, in order to describe the non-Gaussian features of the FPT distributions close to their maximum, we compare them to a Gaussian approximation and to an Edgeworth expansion, which parametrises the first deviations from Gaussian statistics. In this appendix, we recall how the Edgeworth expansion~\cite{Edgeworth:1905,Wallace1958} is constructed.

Let us consider a random variable $X$ with vanishing mean and unit variance, and let $\lbrace X_1, X_2, \cdots X_n\rbrace$ be $n$ independent copies of $X$. We want to determine the distribution function of the normalised sum
\bea
\label{eq:x:def}
x = \frac{1}{\sqrt{n}}\sum_{i=1}^n X_n\, .
\eea 
Since the $X_i$ are independent, $x$ also has a vanishing mean and unit variance.
By virtue of the central-limit theorem, we know that when $n\to\infty$, $x$ becomes a Gaussian random variable, with a vanishing mean and unit variance. This can be seen as the leading-order, Gaussian approximation. Our goal is to go beyond that leading-order result and describe the first non-Gaussian corrections when $n$ is large but finite. 

To that end, we introduce the characteristic function of $x$,
\bea
\chi(t,x) &= \left\langle \exp\left( it x\right) \right\rangle
= \left\langle \exp\left( \frac{it}{\sqrt{n}}\sum_{i=1}^n X_i\right) \right\rangle\\
&=\Pi_{i=1}^n \left\langle \exp\left( \frac{it}{\sqrt{n}}X_i\right)
\right\rangle
=\chi^n\left(\frac{t}{\sqrt{n}},X\right)\, ,
\eea
where we have used that the $X_i$ are independent, and where we recognise $\chi(t,X)$, the characteristic function of the $X$ variable. 

Let us then Taylor expand $\chi(t/{\sqrt{n}},X)$ in the limit where $n$ is large:
\bea
\chi\left(\frac{t}{\sqrt{n}},X\right) &= \left\langle \exp\left(\frac{it}{\sqrt{n}}X\right)\right\rangle\\
&=1
- \frac{t^2}{2n} -i \frac{t^3}{6n\sqrt{n}}\left\langle X^3 \right\rangle + \frac{t^4}{24n^2}\left\langle X^4 \right\rangle+\mathcal{O}\left(\frac{1}{n^2\sqrt{n}}\right),
\eea 
where we have used that $\langle X \rangle =0 $ and $\langle X^2\rangle =1$.
Taking this expression to the $n^\mathrm{th}$ power and further expanding in $n$ leads to
\bea
\chi(t,x) = \ee^{-\frac{t^2}{2}}\left[1-\frac{it^3}{6\sqrt{n}} \left\langle X^3\right\rangle +\frac{t^4}{24n}\left(\left\langle X^4\right\rangle-3\right)-\frac{t^6}{72n}\left\langle X^3\right\rangle^2\right]+\mathcal{O}\left(\frac{1}{n\sqrt{n}}\right)\, .
\eea
This can be cast in terms of the moments of $x$. Since $\langle X_i\rangle=0$ and $\langle X_i X_j\rangle = \delta_{ij}$, \Eq{eq:x:def} leads to $\langle x\rangle =0$, $\langle x^2\rangle =1$, $\langle x^3\rangle = \langle X^3\rangle /\sqrt{n}$ and $\langle x^4\rangle = 3+(\langle X^4\rangle -3) /n$. One thus has
\bea
\chi(t,x) = \ee^{-\frac{t^2}{2}}\left[1-\frac{it^3}{6} \left\langle x^3\right\rangle +\frac{t^4}{24}\left(\left\langle x^4\right\rangle-3\right)-\frac{t^6}{72}\left\langle x^3\right\rangle^2\right]+\mathcal{O}\left(\frac{1}{n\sqrt{n}}\right)\, .
\eea
The PDF can then be obtained by Fourier transforming this expression, see \Eq{eq:PDF:from:chi}, and this leads to 
\begin{equation}
    \label{eq_edgeworth_expansion:interm}
    P(x) \simeq \frac{e^{-x^2/2}}{\sqrt{2\pi}}\left[ 1+\frac{\langle x^3\rangle}{6}He_3 (x) +\frac{(\langle x^4\rangle -3)}{24}He_4(x)  + \frac{\langle x^3 \rangle^2}{72}He_6(x)  \right] \, ,
\end{equation}
where
\begin{align}
    \label{eq_appendix_hermite_polynomials}
     He_3(x) & = x^2-1 \, , \\
     He_4(x) & = x^4-6x^2+3 \, , \\
     He_6(x) & = x^6 -15x^4+45x^2-15 \, ,
\end{align}
are the $3^{\mathrm{rd}}$, $4^{\mathrm{th}}$ and $6^\mathrm{th}$ Hermite polynomials respectively. This expression can be readily generalised to the case where $x$ has a non-trivial mean $\mu$ and standard deviation $\sigma$, by applying the above to $\tilde{x}\equiv (x-\mu)/\sigma$. One thus finds
\begin{equation}
    \label{eq_edgeworth_expansion}
    P(x) \simeq \frac{e^{-\frac{(x-\mu)^2}{2\sigma^2}}}{\sqrt{2\pi\sigma^2}}\left[ 1+\frac{\gamma}{6}He_3 \left(\frac{x-\mu}{\sigma}\right) +\frac{\kappa}{24}He_4\left(\frac{x-\mu}{\sigma}\right)  + \frac{\gamma^2}{72}He_6\left(\frac{x-\mu}{\sigma}\right)  \right] \, ,
\end{equation}
where $\gamma$ and $\kappa$ are called ``skewness'' and ``excess kurtosis'' respectively, and are given by
\bea
\gamma=\frac{\langle \left(x-\mu\right)^3\rangle}{\sigma^3}
\quad\text{and}\quad 
\kappa=\frac{\langle \left(x-\mu\right)^4\rangle}{\sigma^4}-3\, .
\eea 

For a Gaussian random variable, $\gamma$ and $\kappa$ vanish, and one recovers indeed that $P(x)$ is a Gaussian function. Otherwise, the above expression allows one to parametrise small deviations from Gaussianity. It involves the first few moments of the PDF, which in our case can be computed exactly since the characteristic function is known analytically, using the formula
\bea
\langle \N^n\rangle =i^{-n} \left.\frac{\partial^n}{\partial t^n}\chi(t,\phi)\right\vert_{t=0}\, .
\eea 

We find that the Edgeworth expansion provides a reliable approximation in the drift-dominated regime close to the maximum of the distribution, and is therefore a useful comparison for the \code package in the near tail of $P(\N)$. However, it breaks down in the presence of exponential tails, where higher than $4^{\mathrm{th}}$ order moments yield significant contributions.

\bibliographystyle{JHEP}
\bibliography{main.bib}
\end{document}